\documentstyle[preprint,aps]{revtex}   
\renewcommand{\today}{8th June 2000}

\newcommand{\nc}{\newcommand}
\nc{\be}{\begin{equation}}
\nc{\ee}{\end{equation}}
\nc{\bea}{\begin{eqnarray}}
\nc{\eea}{\end{eqnarray}}
\nc{\beas}{\begin{eqnarray*}}
\nc{\eeas}{\end{eqnarray*}}
\nc{\noi}{\noindent}
\nc{\sD}{\not \! \! D}
\nc{\s}[1]{\not \! #1}
\nc{\non}{\nonumber}
\nc{\bb}{\bibitem}
\nc{\lf}{\left}
\nc{\ri}{\right}
\nc{\mb}[1]{\makebox[#1]{}}
\nc{\pa}{\partial}
\nc{\sA}{\not \! \! A}
\nc{\newsec}[1]{\section{#1}\mb{0.5cm}}
\nc{\h}{\frac{1}{2}}
\nc{\ra}{\rightarrow}
\nc{\la}{\leftarrow}
\nc{\ep}{$e^+e^-\ra\pi^+\pi^-\;$}
\nc{\emuon}{$e^+e^-\ra\mu^+\mu^-\;$}
\nc{\epp}{$e^+e^-\ra\pi^+\pi^0\pi^-\;$}
\nc{\elec}{$e^+e^-\ra\gamma^*\ra e^+e^-\;$}
\def\mathunderaccent#1{\let\theaccent#1\mathpalette\putaccentunder}
\def\putaccentunder#1#2{\oalign{$#1#2$\crcr\hidewidth
\vbox to.2ex{\hbox{$#1\theaccent{}$}\vss}\hidewidth}}

\nc{\ti}{\mathunderaccent\tilde}
\nc{\M}{{\cal M}}
\nc{\rw}{$\rho\!-\!\omega\;$}

\begin{document}
\tightenlines    
\draft           
\preprint{\vbox{To appear in Eur. Phys. J. {\bf C}  \\
                                       \null \hfill LPNHE 00--01 \\
                                       \null \hfill SLAC--PUB--8407 \\
                                       \null\hfill nucl-th/0004005 }}
\title{An Effective Approach to VMD at One Loop Order and \\
the Departures from Ideal Mixing for Vector Mesons}
\author{M.~Benayoun, L. DelBuono,  
        Ph. Leruste} 
\address{LPNHE des Universit\'es Paris VI et VII--IN2P3, Paris,
         France}
\author{H.B. O'Connell\footnote{Supported by the US Department of Energy
under contract DE--AC03--76SF00515}}
\address{Stanford Linear Accelerator Center, Stanford University,
         Stanford CA 94309, USA}
\date{Released April 4th, 2000, Revised \today}
\maketitle
\begin{abstract}
We examine the mechanisms producing  departures
from ideal mixing for vector mesons within  the 
context of the Hidden Local Symmetry (HLS) model.
We show that  kaon loop transitions between the ideal 
combinations of the $\omega$ and $\phi$ mesons
necessitate a field transformation in order to get
the mass eigenstates.  It  is shown that
this transformation is close to a rotation for
processes involving, like meson decays,
 on--shell $\omega$ and $\phi$ mesons.
The HLS model predicts  a momentum    dependent, 
slowly varying mixing angle between the ideal states. We
examine numerically the consequences of this
for radiative and leptonic decays of light mesons.
The mean $\omega-\phi$  mixing angle is found 
smaller than its ideal value; this is  exhibited separately in 
radiative and in leptonic decays. Effects of nonet symmetry breaking 
in the vector sector  are compared to those produced by 
the field rotation implied by the HLS model.
\end{abstract}

\newpage

\pagenumbering{arabic}
\section{Introduction}

The Hidden Local Symmetry (HLS) Model in both its non--anomalous
\cite{HLS} and anomalous (FKTUY) sectors \cite{FKTUY} is 
a powerful tool for analyzing experimental data, by
providing a clear framework with the fewest possible number of free
parameters. For instance, it allows a 3--parameter description of the
$I=1$ pion form factor;  this gives a statistically optimal description in
an energy interval running from threshold to the $\phi$ mass. This has
been shown by Ref.~\cite{ben0} in analyzing the world data set for $e^+e^-
\ra \pi^+ \pi^-$ annihilation collected in Ref.~\cite{barkov}. The
exercise has been repeated as successfully with the data set recently
collected by the CMD--2 Collaboration on the VEPP--2M collider at
Novosibirsk \cite{akhmetshin}.

However, in order to go beyond while staying within the 
framework defined by the HLS model and its  anomalous sector, one needs to define a 
consistent scheme of symmetry breakings. Without SU(3) breaking, the HLS model cannot
successfully describe the kaon form factors; without nonet symmetry
breaking in the pseudoscalar (PS) sector, it cannot be used reliably
to describe radiative decays of light mesons. The BKY
mechanism \cite{BKY,heath} is a consistent way to introduce SU(3) breaking
in both the vector (V) and PS sectors.  It has been shown
recently~\cite{chpt} that the BKY SU(3) breaking in the PS sector is in
perfect agreement with all accessible predictions of Chiral Perturbation
Theory~\cite{leutw,leutwb,feldmann} at first order in the breaking
parameters.  In order to reach this conclusion, the needed ingredients
were only the BKY breaking in the PS sector (referred to hereafter as
$X_A$ breaking), the kinetic energy term of the non--anomalous HLS
Lagrangian and the $P\gamma \gamma$ Lagrangian of Wess, Zumino and Witten
\cite{WZW}. Thus, this part is on secure grounds.

When dealing with PS mesons, the question of nonet symmetry breaking (NSB)
cannot be avoided, as clear from Refs.~\cite{veneziano,shore} for instance. 
It was already introduced in the physics of single photon radiative decays (12
modes) by O'Donnell long ago \cite{odonnel}, relying  basically only
on group theoretical considerations, but outside the
context of effective Lagrangians. Thus, Ref.~\cite{odonnel} already proved
that nonet symmetry breaking  can be dealt with simply in both the V and PS sectors.  
However, the way to include  nonet symmetry breaking of both kinds consistently within 
a Lagrangian (HLS for instance) is not necessarily a trivial matter.

Recently, we proposed \cite{rad} a model to describe, 
 within the HLS framework, all radiative decays of light
mesons ($VP\gamma$ or $P \gamma \gamma$ processes) and their leptonic
decays ($\rho^0/\omega /\phi \to e^+e^-$ processes). Ref.~\cite{rad}
included SU(3) breaking in the V and PS sectors in the manner of BKY
\cite{BKY,heath} and proposed a way to include simultaneously NSB in the
PS sector. Slightly later, Ref.~\cite{chpt} showed that this mixed NSB and
$X_A$ breaking scheme can be derived rigorously by supplementing the
non--anomalous HLS Lagrangian with a piece coming from the ChPT
Lagrangian, ${\cal L}_1$ \cite{leutw,leutwb}.

In this context, the BKY SU(3) breaking in the vector sector (referred to as
$X_V$) acts only in leptonic decays. As the data description was already
quite satisfactory \cite{rad,chpt}, a possible breakdown of  nonet symmetry
in the vector sector was not examined; indeed, it could be guessed that this  would be
beyond the accuracy of the available data, at least if
vector NSB shows up only at the coupling constant level, as
inferred in Ref.~\cite{odonnel}.

Except for SU(3) breaking effects, the model in Ref.~\cite{rad}
recovers the structure exhibited long ago by O'Donnell \cite{odonnel}.
Fundamental parameters common to Refs.~\cite{rad,odonnel} are two
mixing angles. $\delta_P$ has a clear origin \cite{DGH} and
is essentially generated by SU(3) breaking in the PS sector. Ref.~\cite{chpt}
has shown that the traditional single PS mixing angle $\theta_P$ is algebraically
related with the ChPT  angle \cite{leutw,leutwb} and that $\theta_8 \simeq 2 \theta_P$.

Another fundamental ingredient of this model is an angle, $\delta_V$,
which describes departures of the $\omega/\phi$ system from ideal mixing
by bringing some strange quark content into the $\omega$ meson and,
conversely, some non--strange component into the $\phi$ meson. This angle
was already a basic ingredient in the axiomatic model of O'Donnell
\cite{odonnel}. Its origin within effective Lagrangians has not yet been
worked out thoroughly, even if it is clearly connected with $\omega
\leftrightarrow \phi$ transitions \cite{heath,tony} inherent to VMD models
like the HLS model \cite{HLS}.

Therefore, it looks relevant to study in some details how $\omega
\leftrightarrow \phi$ transitions in the HLS model generate mixing of the
ideal combinations into the physical $\omega$ and $\phi$; whether this
correspondence is stricly speaking a rotation, as commonly assumed
\cite{odonnel,rad}, or a more complicated transformation, is also an
interesting issue. It is also interesting to explore the origin
of such a transformation and see whether this could be attributable to
some sort of vector NSB.

We address these topics using the method of effective Lagrangians, as
higher orders in the HLS model would meet the problem of its
renormalizability which is not our concern.  At each step, however, we 
have examined analyticity properties in connection
with what can be inferred from the Analytic S--matrix Theory \cite{ELOP}.
Finally, we shall comment on the way VMD has to be broken \cite{rad} 
in order to account effectively for the observed ratio of yields 
$[K^{*0} \ra K^0 \gamma]/[K^{*\pm} \ra K^\pm \gamma] \simeq 2$
which is naturally possible within other contexts\cite{odonnel,GM}.

The outline of the paper is as follows. In Section \ref{two} we briefly
remind  the reader of the BKY SU(3) breaking scheme we use. In
Section \ref{three}, we define the effective Lagrangian piece which
accounts for loop corrections; this is done essentially by
relying on the results provided by the Schwinger--Dyson equation in the
$\omega/\phi$ sector. In Section \ref{four}, we show that the
diagonalization of the vector mass term at one--loop order is well
approximated by a simple rotation, albeit momentum dependent, 
for on--shell $\omega$ and $\phi$ mesons; we
also discuss somewhat  the  concept of mass--shell for vector mesons. 
 In Section \ref{five}, we remind how NSB in
the PS sector is generated; we also discuss some variants of a possible
vector NSB which are numerically studied. In Section
\ref{six}, we recall a Lagrangian for the anomalous sector, equivalent to
FKTUY when no vector NSB exists; however, it allows one to recover 
exactly the O'Donnell model in its full generality. Section \ref{fiveb} recalls the
breaking procedure which permits  to include both $K^*$ decay modes
within the set of successfully fitted radiative modes; we argue that this
might imply a renormalization of the vector fields presently missing in
the BKY breaking scheme.

Section \ref{seven} is devoted to the numerical analysis of our model of
radiative and leptonic decays under various conditions, noticeably with
the momentum--dependent $\delta_V$ which can be inferred
from the HLS model. Effects of vector meson self--energies on vector meson
masses are briefly exemplified in Section \ref{eight}. The fit results
obtained are commented on in Section \ref{nine}, while Section \ref{ten}
is devoted to conclusions. Finally, analytic
expressions for some loops, vector meson self--energies and transition
amplitudes, coupling constants in the general case, are gathered in the
Appendices in order to make  easier reading of the main text.

\section{SU(3) Breaking of the Non--Anomalous HLS Lagrangian}
\label{two}

Let us denote by $P$ and $V$ the bare pseudoscalar and vector field
matrices under the assumption of nonet symmetry.  With the convention we
use in this paper, they can be found in Ref.~\cite{rad} (see Eqs.~(6) and
(9) there; notice that our ideal $\phi$ field, denoted $\phi_I$, is
$-s\overline{s}$). Let us denote by $A$ the electromagnetic field and let
$Q={\rm Diag}(2/3,-1/3,-1/3)$ be the quark charge matrix.

Let us also denote by $X_A={\rm Diag}(1,1,\ell_A)$ and $X_V={\rm
Diag}(1,1,\sqrt{\ell_V})$ the SU(3)  breaking matrices in, respectively,
the PS and V sectors following from the BKY mechanism \cite{BKY,heath}.
The unbroken limit is obtained by stating $\ell_A=\ell_V=1$. Then, the
SU(3) broken HLS Lagrangian can be written \cite{BKY,heath}

\be
{\cal L} = {\cal L}_A + a{\cal L}_V
\label{HLS1}
\ee
\noindent where $a$ is the standard HLS parameter \cite{HLS}.
Following the breaking scheme proposed in Ref.~\cite{heath}
(see Section III.D therein), we have
\begin{equation}
\left \{
\begin{array}{lll}
{\cal L}_A&=&{\rm Tr}[\pa PX_A\pa PX_A+2ie(PQA-AQP)X_A\pa PX_A]\\[0.5cm]
\label{NS0}\\
{\cal L}_V&=&{\rm Tr}[f_\pi^2((g V-eQA)X_V)^2
+i(g V-eQA)X_V(\pa PP-P\pa P)X_V]
\end{array}
\right .
\label{HLS2}
\end{equation}
where $f_\pi$ is the pion decay constant (92.42 MeV \cite{PDG98}), $g$ is
the universal coupling strength of vector mesons and $e$ is the unit
electric charge. The corresponding Lagrangian can be found
expanded{\footnote{ In Ref.~\cite{heath}, the fields named $\omega$ and
$\phi$ correspond to what we call here $\omega_I$ and $-\phi_I$.
}} in Ref.~\cite{heath} as Eq. (A5).

The properties of the kaon form factor at $s=0$ imply \cite{BKY,heath}
that{\footnote{We might use indifferently some notations related
to each other \cite{rad,chpt} $\ell_A=z=1/Z$ for backward compatibility.}}
$\ell_A\equiv [f_K/f_\pi]^2$. This is quite interesting as it implies
that SU(3) breaking in the PS sector does not result in additional free
parameters ($\ell_A=1.5)$; this has been checked with experimental data in
Ref.~\cite{rad}. Likewise, $\ell_V$ is connected with the ratio of the
vector meson Higgs--Kibble (HK) masses \cite{BKY,heath}
$m_\phi^2/m_\omega^2=m_\phi^2/m_\rho^2=\ell_V$; practically, this relation
is, however, less interesting than the previous one since it implies 
that the Lagrangian (HK) masses can be extracted reliably  from data. 

In order to restore the kinetic energy term of the PS mesons to its
canonical form, the $X_A$ breaking results into a field
renormalization\cite{BKY,heath}

\be
P_{ren}= X_A^{1/2}PX_A^{1/2}
\label{HLS3}
\ee

On the other hand, the $X_V$ breaking \cite{BKY,heath} results in a mass
breaking in the vector meson sector, but not in renormalization{\footnote{
The Yang--Mills kinetic term is added to the HLS Lagrangian but does not follow
from its construction.  On the other hand, we shall see in Section \ref{fiveb}
that data might give a hint in favor of a renormalization of the vector
fields. Any renormalization of these would obviously break the relation between
$\ell_V$ and the vector meson masses. }} of the vector fields. After this
breaking, the mass term of neutral vector mesons becomes \cite{heath}

\begin{equation}
{\cal M}= \displaystyle \frac{a f_\pi^2g^2}{2}[\rho^2 + \omega_I^2+\ell_V\phi_I^2]
\label{HLS4}
\end{equation}

We define for further use $m_\rho^2=a g^2 f_\pi^2$. It should be clearly
stated here once more that these masses might have little to do with the
observed peak positions as reported in Ref.~\cite{PDG98}. This is due
to the difficulty of estimating unambiguously effects of the real part of 
vector meson self--energies. This clearly follows from the form factor studies 
in Refs.~\cite{klingl,shin} and will also be illustrated below. On the other
hand, the definition itself of vector meson masses is not unambiguous, as
exemplified in
Refs.~\cite{tony,Gardner:1998ie,Eidelman:1998jj,pennington}; this question
is discussed in Sections  \ref{three} and \ref{eight}.

\section{An Analytic Approach to VMD at One--Loop Order}
\label{three}

As soon as one considers non--leading effects produced by the various
SU(3) breaking procedures, consistency implies to account also for other
non--leading effects which proceed from the HLS Lagrangian itself,
irrespective of any breaking procedure. In this Section, we examine the
contributions originating from loops and show how they can be incorporated
{\it effectively} into the HLS Lagrangian.  The purpose is to define a
coherent framework, able to allow  a phenomenological study of experimental
data.

First of all, even if non--leading, loops produce observable effects in
measurable (and measured) quantities.  An illustrative example is
the pion form factor in the timelike region; in this case, the shape 
observed is essentially an
effect of dressing the $\rho^0$ meson propagator, noticeably an effect of
the pion loop. As clear from Refs.~\cite{ben0,klingl}, the detailed
structure of the pion form factor is perfectly reproduced by the pion
loop, and the $\pi \pi$ phase shift too. So, a coherent phenomenological
handling of data implies to account properly for these leading loop
effects. If pion form factor studies \cite{ben0,klingl} give a clear hint
that loop effects in vector meson self--energies play an essential role,
loop corrections to the $\rho \pi \pi$ vertex do not show up. In the
following such kind of one--loop effects are neglected.

In the present work, we focus on loops generated by the non--anomalous 
HLS Lagrangian. Thus, we do not consider  loops generated by 
the anomalous VVP Lagrangian ($VP$ or $P \gamma$) or double loop effects produced 
by the VPPP anomalous Lagrangian. They do not change qualitatively 
the picture; however, their possible effects in  the realm of state mixing and meson decays
will be commented on at the relevant places.  

\subsection{Effects of Loops on Vector Meson Masses}

Let us neglect all electromagnetic contributions from the HLS
Lagrangian{\footnote{ Within the VMD framework, the
photon has a special status. Indeed, the Lagrangian expressed in terms of
physical fields has a quadratic term which contains mixed terms $\gamma
\rho^0$, $\gamma \omega$, $\gamma \phi$.  It is the essence of VMD to keep
these transition terms. For a simultaneous handling of
$\rho^0$ and photon within the HLS model, see Ref.~\cite{tony}.}} and work
at one--loop order; we also denote from now on $\rho$ the $\rho^0$ field
when there is no ambiguity.  Amputated from its Lorentz tensor part, the
dressed $\rho$ propagator $D(s)$ is given by the Schwinger--Dyson equation
\cite{tony};  we have at one loop ($g^2$) order
\be
D^{-1}(s)=D_0^{-1}(s)-\Pi_{\rho \rho}(s)
\label{tony1} 
\ee
where the inverse bare propagator is $D_0^{-1}(s)=s-m_\rho^2$ and
$\Pi_{\rho \rho}(s)$ is the $\rho$ meson self--energy which contains
contributions from the pion and both kaon loops \cite{tony}.  The $\rho$
self--energy is given in Appendix \ref{BB} and can be explicitly
constructed using formula from Appendix \ref{AA}. This can be {\it
effectively} obtained from the HLS Lagrangian by adding a piece
$\Pi_{\rho \rho}(s)\rho^2/2$ which turns out to modify the $\rho$ mass
term coefficient to $[m_\rho^2+\Pi_{\rho \rho}(s)]/2$. This (effectively)
modified Lagrangian ${\cal L}'(s)$ gives automatically the dressed $\rho$
propagator requested by phenomenology  and coincides with the
solution to the Schwinger--Dyson equation. Below the two--pion threshold
we have $ {\cal L}^{'}(s)={\cal L}^{'\dagger}(s)$, for any real $s$.
Above, the hermiticity condition is naturally extended to ${\cal
L}^{'}(s)={\cal L}^{'\dagger}(s^*)$, where the symbol $*$ denotes
complex conjugation; this property, known as hermitian analyticity
\cite{ELOP,philippe}, is indeed fulfilled by ${\cal L}^{'}(s)$.

Therefore, using ${\cal L}^{'}(s)$ turns out to include directly one--loop
corrections into the HLS Lagrangian in a way consistent with the
Schwinger--Dyson equation and in agreement with the analyticity
assumption.  The HLS Lagrangian can be {\it effectively} modified
correspondingly in order to include also one--loop effects associated with
charged $\rho$'s and with all $K^{*}$ fields.  One should stress that
these additional pieces {\it a priori} depend on the invariant mass
squared $s$; they are given in Appendix \ref{BB} and can be constructed
using the results in Appendices \ref{AA} and \ref{AA2}.

As the considered fields are effective fields, this $s$ dependence is not
really an issue and should only reveal non--local effects due to the fact
that vector fields are functions of more fundamental ones (quark and gluon
degrees of freedom). Additionally, this $s$--dependence is a natural
feature at hadron level; it tells that the mass of a resonance does
not fulfill $s-m_V^2=0$ any longer, but $s-m_V^2-\Pi_{VV}(s)=0$. In the
case of the $\rho$ and $K^*$ mesons, this proves that the physical mass is
shifted into the complex $s$--plane and that the connection between the HK
mass $m_V^2$ and the real part of the pole is somewhat lost (see
Ref.~\cite{tony} for the $\rho$ meson).

\subsection{The special case of the $\omega$ and $\phi$ mesons}

In the $\omega_I/\phi_I$ sector, the situation is slightly more complicated.
In this case, the  Schwinger--Dyson equation is the $2\times 2$ 
matrix equation 

\be
D^{-1}(s)=
\left(
\begin{array}{ll}
s-m_{\omega_I}^2 &~~~0\\[0.5cm]
~~~0 & s-m_{\phi_I}^2
\end{array}
\right)
-
\left(
\begin{array}{ll}
\Pi_{\omega_I\omega_I}(s) & \Pi_{\omega_I\phi_I}(s)\\[0.5cm]
\Pi_{\phi_I\omega_I}(s)     &\Pi_{\phi_I\phi_I}(s)  
\end{array}
\right)
\label{tony2}
\ee 
which exhibits the dressing of the mass terms as for $\rho$ and $K^*$
mesons, but also a non--diagonal term.  The pieces generating these loops
at order $g^2$ are given by the following $V-PS$ interaction term from the
${\cal L}_V$ part of the HLS Lagrangian \cite{heath}

\begin{equation}
\begin{array}{lll}
{\cal T}_1= & \displaystyle -\frac{iag}{4} Z[\omega_I +\sqrt{2} \ell_V \phi_I]
\left [ K^+ \stackrel{\leftrightarrow}{\pa} K^- 
+ K^0\stackrel{\leftrightarrow}{\pa}\overline{K}^0 \right ]\\[0.3cm]
~&-\displaystyle\frac{iag}{4}\rho
[Z \left (  K^+ \stackrel{\leftrightarrow}{\pa} K^- - 
K^0\stackrel{\leftrightarrow}{\pa}\overline{K}^0 \right)
+2 \pi^+\stackrel{\leftrightarrow}{\pa}\pi^-]
\end{array}
\label{HLS5}
\end{equation}
where we have renormalized the kaon fields according to Eq. (\ref{HLS3}) above
($Z=[f_\pi/f_K]^2$).

At first non--leading order, this term generates $K \overline{K}$ and
$\pi^+\pi^-$ loops. Both $K \overline{K}$ loops generate the $\omega_I$
and $\phi_I$ self--energies $\Pi_{\omega_I\omega_I}(s)$ and
$\Pi_{\phi_I\phi_I}(s)$. They also produce $\omega_I \leftrightarrow
\phi_I$ transitions with a non--zero amplitude $\Pi_{\omega_I\phi_I}(s)$.  
The $\rho$ self--energy $\Pi_{\rho \rho}(s)$ receives contributions from
the pion loop and from both kaon loops, as stated above.  However, as
SU(2) symmetry is assumed, there is no $\rho \leftrightarrow \omega_I$ or
$\rho \leftrightarrow \phi_I$ transition loops. Indeed, in these cases,
charged and neutral kaon loops exactly compensate \cite{tony}. This is why
we have decoupled the $\rho^0$ channel while writing Eq. (\ref{tony2}).

In contrast with $\rho$ and $K^*$ fields which remain mass eigenstates
with only one--loop modified masses, Eq. (\ref{tony2}) shows that
$\omega_I$ and $\phi_I$ are no longer mass eigenstates at one--loop order.  
The physical masses are (complex) solutions of Det$[D^{-1}(s)]=0$. This
equation can also be written $[s-\lambda_\omega(s)][s-\lambda_\phi(s)]=0$,
in terms of the eigenvalues of
\be
M^2=
\left(
\begin{array}{ll}
\Pi_{\omega_I\omega_I}(s)+m_{\omega_I}^2 & ~~~\Pi_{\omega_I\phi_I}(s)\\[0.5cm]
~~~\Pi_{\phi_I\omega_I}(s)     &\Pi_{\phi_I\phi_I}(s) +m_{\phi_I}^2  
\end{array}
\right)
\label{tony3}
\ee

They  depart from the HK masses ($\lambda_\omega=m_{\omega_I}^2$ and
$\lambda_\phi=m_{\phi_I}^2$) because of loops and become running.
The  momentum--dependent 
analytic matrix{\footnote{The matrix $G(\delta_V)$ can be handled
as if it were actually an orthogonal matrix, even  
for  complex values of $s$ or for real $s$ above $4 m_K^2$.
Here and  in the following, trigonometric functions should
be understood as their underlying exponential expressions.
Then, all usual trigonometric relations apply, even for complex arguments, 
reminding that $\cos{iu}=\cosh{u}$, $\sin{iu}=i\sinh{u}$,
etc\ldots  }}  
\begin{equation}
G(\delta_V)=
\left(
   \begin{array}{llll}
        ~~\cos{\delta_V} & \sin{\delta_V}   \\[0.5cm]
	-\sin{\delta_V} & \cos{\delta_V}  \\
    \end{array}
\right)
\label{tony4}
\end{equation}
with 
\begin{equation}
\tan{2\delta_V}(s) = \displaystyle\frac{2 \Pi_{\omega_I\phi_I}(s)}
{m_\rho^2(1-\ell_V)+(\Pi_{\omega_I\omega_I}(s)-\Pi_{\phi_I \phi_I}(s)) }
\label{HLS8}
\end{equation}
diagonalizes Eq. (\ref{tony2}). The first term in the denominator is
actually $m_{\omega_I}^2-m_{\phi_I}^2$, in terms of the HK masses coming
out from the BKY broken HLS Lagrangian \cite{heath}. Indeed, we have

\begin{equation}
D_G^{-1}=  [GDG^{-1}]^{-1}= s I - GM^2G^{-1}= sI-
\left(
  \begin{array}{ll}
  \lambda_\omega(s) & ~~~0 \\[0.5cm]
 ~~~0& \lambda_\phi(s)
    \end{array}
\right)
\label{tony5}
\end{equation}
(we have always $G^{-1}(\delta_V)=G(-\delta_V)$ as if $G$ were always an actual
orthogonal matrix), and the eigenvalues

\begin{equation}
  \begin{array}{ll}
\lambda_\omega(s)=  [m_{\omega_I}^2+\Pi_{\omega_I\omega_I}(s)] \cos^2{\delta_V}+
[m_{\phi_I}^2+\Pi_{\phi_I\phi_I}(s)]\sin^2{\delta_V}
  +\Pi_{\omega_I\phi_I}(s) \sin{2\delta_V}\\[0.5cm]
\lambda_\phi(s)=  [m_{\phi_I}^2+\Pi_{\phi_I\phi_I}(s)]\cos^2{\delta_V}+
[m_{\omega_I}^2+\Pi_{\omega_I\omega_I}(s)] \sin^2{\delta_V}
  -\Pi_{\omega_I\phi_I}(s) \sin{2\delta_V}
 \end{array}  
\label{tony6}
\end{equation}
are the running squared masses associated with the physical eigenstates of
the HLS Lagrangian at one--loop order. In scattering processes the
corresponding dressed propagators are $D^{-1}_\omega(s)=s -
\lambda_\omega(s)$ and $D^{-1}_\phi(s)=s-\lambda_\phi(s)$.

The ``trigonometric'' functions entering Eqs. (\ref{tony6}), can be expressed
easily in terms of the right--hand side of Eq. (\ref{HLS8}) using
\begin{equation}
  \begin{array}{ll}
\cos{2 \delta_V}(s)= \displaystyle \frac{1}{(1+\tan^2{2\delta_V}(s))^{1/2}}~~~, 
&\sin{2 \delta_V}(s)= \displaystyle \frac{\tan{2\delta_V}(s)}{(1+\tan^2{2\delta_V}(s))^{1/2}}
 \end{array}  
\label{tony7}
\end{equation}

A detailed study of the singularities of the eigenvalues and eigenvectors
as real--analytic functions
is beyond the scope of this paper. Let us only mention that 
renormalization conditions (see below) on self--energies might have to
compensate (possible) simple zeros  of $1+\tan^2{2\delta_V}(s)$ located in the 
physical sheet  of the scattering amplitudes. On the other hand,
Det$[D^{-1}(s)]$ is  tightly connected with the $D$ function of the former $N/D$
formalism \cite{ELOP,philippe,chew} and has certainly all desired analyticity 
properties.

The {\it physical} $\omega$ mass is solution of $s-\lambda_\omega(s)=0$,
while the {\it physical} $\phi$ mass is separately solution of
$s-\lambda_\phi(s)=0$. Physical $\omega$ and $\phi$ fields correspond to
these mass eigenstates; they are obviously associated with the
eigenvectors of the matrix $M^2$ of Eq. (\ref{tony3}). Each physical field can be identified by 
inspecting the behavior of the eigensolutions at $\delta_V \ra 0$. This 
identification is meaningful as the effective value of $\delta_V$ is small, 
a few  degrees only \cite{rad}.

\subsection{Two Regimes For The $\omega-\phi$ Mixing}

It is easy to check that all features described in the two previous
Subsections can be obtained by adding the following {\it effective} piece
to the HLS Lagrangian

\begin{equation}
\begin{array}{lll}
{\cal L}_{{\rm loops}}(s)= &\displaystyle \frac{1}{2}
[\Pi_{\rho \rho}(s)\rho^2+ \Pi_{\omega_I\omega_I}(s) ~\omega_I^2+\Pi_{\phi_I\phi_I}(s)~\phi_I^2+ 
2\Pi_{\omega_I\phi_I}(s)~\omega_I  \phi_I]\\[0.3cm]
~~& + \displaystyle [\Pi_{\rho \rho}(s) \rho^+\rho^- +\Pi_{K^* K^*}(s) (K^{*+}K^{*-}+K^{*0}\overline{K}^{*0})]
\end{array}
\label{loop1}
\end{equation}

All needed loops and self--energies are gathered in Appendices
\ref{AA}--\ref{BB}; couplings constants can be identified by examining 
the Lagrangian in Eq. (A5) of Ref.~\cite{heath} and are not listed here.
They should be modified according to field renormalization procedures.
Therefore, the use of ${\cal L}_{{\rm loops}}(s)$ turns out to account
directly for one--loop effects into the propagators, in a way consistent
with the Schwinger--Dyson equation. This Lagrangian piece fulfills
the hermitian analyticity condition  \cite{ELOP}
${\cal L}_{{\rm loops}}(s)={\cal L}_{{\rm loops}}^\dagger(s^*)$.

In the approximation where electromagnetic contributions and anomalous
Lagrangian terms \cite{FKTUY} are neglected, all loops involved in the
$\omega/\phi$ sector are basically the $K\overline{K}$ loop only. Let us
denote by $\Pi(s)$ the $K\overline{K}$ loop amputated from the coupling
constants to vector mesons and state $\Pi_{VV'}=
g_{VK\overline{K}}g_{V'K\overline{K}}\Pi(s)$ for any pair of vector mesons
$(V,V')$ coupling to kaon pairs.

There are clearly two regimes for the $\omega-\phi$ mixing, depending on
the value of $s$ involved. As clear from Appendix \ref{AA}, $\Pi(s)$ is
real for real $s$ such that $s\leq 4 m_K^2$ and acquires an imaginary part
when $s\ge 4 m_K^2$. The $\omega_I-\phi_I$ transitions affect the dressed
propagators already at one--loop order, as illustrated above. However, the
same transition loops affect also $\omega_I$ and $\phi_I$ as external legs.
The diagonalization of the full mass term

\begin{equation}
{\cal M}(s)= \displaystyle \frac{a f_\pi^2g^2}{2}[\omega_I^2+\ell_V\phi_I^2]+
\frac{1}{2} [\Pi_{\omega_I\omega_I}(s) ~\omega_I^2+\Pi_{\phi_I\phi_I}(s)~\phi_I^2+ 
2\Pi_{\omega_I\phi_I}(s)~\omega_I  \phi_I]
\label{HLS6}
\end{equation}
of the Lagrangian ${\cal L}_{{\rm loop}}+ {\cal L}_{HLS}$, defines
the effective physical fields and also diagonalizes the Schwinger--Dyson equation.
The physical $\omega$ and $\phi$ fields have running squared masses
$\lambda_\omega(s)$ and $\lambda_\phi(s)$ respectively.

For real $s\leq 4 m_K^2$, $G(\delta_V)$ is indeed a (mass--dependent)
rotation while above it is not. However, the transformation $G$ is still
valid and fulfills $G^{-1}(\delta_V)=G(-\delta_V)$. $\delta_V(s)$ can
always be split up into its real and imaginary parts
$\delta_V(s)=\alpha_V(s) +i \beta_V(s)$ which are no longer (separately)
analytic functions; $\alpha_V(s)$ and $\beta_V(s)$ can be
derived from Eq. (\ref{HLS8}).

The mass matrix $M^2$ is rendered diagonal by  physical fields which
are combinations of the ideal states. It is easy to check that these 
effective physical fields are
\begin{equation}
\left(
   \begin{array}{ll}
     \omega \\[0.5cm]
     \phi \\
   \end{array}
\right)
=
\left(
   \begin{array}{llll}
        ~~\cos{\delta_V} & \sin{\delta_V}   \\[0.5cm]
	-\sin{\delta_V} & \cos{\delta_V}  \\
    \end{array}
\right)
\left(
   \begin{array}{ll}
     \omega^I \\[0.5cm]
     \phi^I   \\
   \end{array}
\right)
\label{HLS7}
\end{equation}

With this definition of physical fields, the mass term Eq. (\ref{HLS6}) of
the modified HLS Lagrangian becomes canonical{\footnote{The $s$ dependence
of the fields is understood. One could write this expression in a more 
symmetric way, taking into account that $\omega(s) \equiv \omega^\dagger(s^*)$ and 
$\phi(s) \equiv \phi^\dagger(s^*)$.}}

\begin{equation}
{\cal M}(s)= \frac{1}{2}[\lambda_\omega(s) \omega^2 + \lambda_\phi(s) \phi^2]
\label{HLS6b}
\end{equation}
and fulfills ${\cal M}(s)={\cal M}^\dagger(s^*)$.
It is worth noting  that, in the effective approach we follow,
the physical $\omega$  and $\phi$ fields behave like real--analytic
functions of $s$ and are associated with running masses
which are also real--analytic functions of $s$.

Eq. (\ref{HLS7}) is valid for any value of $s$. For complex $s$ or for real 
$s$ above $4 m_K^2$, this can be written more traditionally

\begin{equation}
\left(
   \begin{array}{ll}
     \omega \\[0.5cm]
     \phi \\
   \end{array}
\right)
=
\left(
   \begin{array}{ll}
~~\cosh{\beta_V} & i\sinh{\beta_V} \\[0.5cm]
-i\sinh{\beta_V} &\cosh{\beta_V}\\
    \end{array}
\right)
\left(
   \begin{array}{ll}
~~\cos{\alpha_V} & \sin{\alpha_V} \\[0.5cm]
-\sin{\alpha_V} &\cos{\alpha_V}\\
    \end{array}
\right)
\left(
   \begin{array}{ll}
     \omega^I \\[0.5cm]
     \phi^I   \\
   \end{array}
\right)
\label{HLS7b}
\end{equation}
in terms of the real and imaginary parts of $\delta_V$. For real $s$  
below $4 m_K^2$, $\beta_V$ vanishes and  $G(\delta_V)$
is a rotation of angle as given by Eq. (\ref{HLS8}). This can also
be written  
\begin{equation}
\tan{2\delta_V}(s) = \displaystyle \frac{2\sqrt{2} a Z^2\ell_V \Pi(s)}
{8f_\pi^2(1-\ell_V)+aZ^2(1-2 \ell_V^2)\Pi(s) }
\label{HLS8b}
\end{equation}
in terms of the basic HLS parameters from the broken Lagrangian.
$\Pi(s)$ is the generic kaon loop already defined and depends on a subtraction
polynomial $P(s)$ (see Appendices \ref{AA} and \ref{BB}) discussed in the 
following Subsection. Departures of the $\omega-\phi$  mixing from a pure rotation 
are exhibited in Eq. (\ref{HLS7b}) and will be discussed below.

\subsection{Renormalization Conditions on Loops}

The self--energies we have defined are each given by a dispersion relation
on the imaginary part of a loop function; they are 
analytic functions of $s$ defined on a 2--sheeted Riemann surface with a
second order branch point at threshold. This implies some analytic
properties for propagators \cite{ELOP,philippe} which are not examined
here; let us only mention that physical meson masses correspond to propagator 
poles in the unphysical sheet of a 2--sheeted Riemann surface onto which amplitudes
are defined{\footnote{ Roughly speaking, poles associated with a resonance come by 
pairs;  specializing to 2--body decay channels as in the HLS model, the pole effectively 
associated with the resonance is located at some $s_R$ in the unphysical sheet close to the 
physical region, while there exists a so--called shadow pole \cite{ELOP} located 
at $s^*_R$ in the same Riemann sheet but generally far from the physical region.
This is not necessarily true if resonance poles are located
close to some threshold.}}.

Integrability conditions on the dispersion integral imply that the
dispersion relation should be at least twice subtracted, which gives rise
to a subtraction polynomial $P(s)$ of (at least) first degree with (at
least) two unknown coefficients which have to be fixed by stating some
renormalization conditions: $ P(s)=d_0+d_1 s$. The minimum number of
subtractions is mandatory in order for the integral to make sense; however,
the {\it actual} number of subtractions can be larger and depends on the
assumed behavior at infinity of the function considered. If one performs
$k$ subtractions, giving rise to a polynomial of degree $k-1$, it is
useful (but not mandatory) to require $\lim_{s\ra 0} [\Pi(s)-P(s)] \simeq
{\cal O}(s^k)$.

It is quite usual for all meson self--energies to fix the constant terms
$d_0$ of the polynomials $P(s)$ to zero, in order to ensure that the
photon remains massless \cite{klingl}. It also plays some role \cite{pichowsky}
for current conservation.  This condition is more stringent
than needed as, for instance, masslessness of the photon implies only that some
combination of the $d_0$'s is zero{\footnote{Each relevant loop gives rise
to a polynomial $P(s)$ and all such polynomials are different from each
other. }}.  However, assuming that the $d_0$ parameter associated with
each possible loop is zero implies that each loop fulfills $\lim_{s \ra 0}
\Pi_{VV'}=0$. This condition has the virtue to ensure that each (running)
vector meson mass coincides with its HK mass value at the chiral point.
This is a strong condition which can be assumed, at least provisionally,
because of its aesthetic character.

On the other hand, the renormalization condition $d_0=0$ is also
appropriate for the transition loop $\Pi_{\omega_I \phi_I}(s)$. Indeed, the
condition $\Pi_{\omega_I \phi_I}(s=0)=0$ allows to derive the amplitudes
\cite{rad,chpt} for $P \gamma \gamma$ processes from the FKTUY Lagrangian
\cite{FKTUY} $PVV$ by making $\delta_V(s=0)=0$. Stated otherwise, internal
$\omega$ or $\phi$ lines computed at $s=0$ coincides with $\omega_I$ or
$\phi_I$ respectively. In the mass region of vector resonances, the condition 
$\delta_V=0$ is no longer fulfilled  for obvious reasons (see Eq. (\ref{HLS8b})).

Therefore, the $s$--dependence of $\delta_V$ is a fundamental property as
it allows to make consistent a zero value for mixing angle at $s=0$ with
non--zero values in the mass region of vector resonances.

Concerning the $\omega/\phi$ mixing which is presently our main concern,
it remains to fix the rest of the polynomial $P(s)$. As there is no clear
and unambiguous statement about the remaining coefficients, it follows
that the phase shift $\delta_V(s)$ is somewhat free, or that its value can
be used as additional renormalization condition.  As explicit in Eq. 
(\ref{HLS7b}), we have assumed $\Pi_{VV'}=g_{VK\overline{K}} g_{V'K\overline{K}} \Pi(s)$; 
this is actually a strong assumption which serves only to lessen the number of
free parameters (or of needed renormalization conditions) in physical
problems, as the subtraction polynomials (not the rest!) might be not
related{\footnote{If indeed related, the proof is anyway not completely
straightforward.}} by simple rescaling by the appropriate product 
of coupling constants. This point is commented on slightly more
in Appendix \ref{BB}.

\section{Vector Meson Mass--Shell and On--Shell $\omega-\phi$ Mixing}
\label{four}

Particle mass--shell is a well defined concept for objects
like PS mesons. For others, like vector mesons, this concept
is somewhat  more embarrassing. Indeed, one can choose
to define it as  zero of the real part of the inverse propagator
of this particle; this is recommended by the Particle Data Group \cite{PDG98}
but not free from ambiguities, depending on coupled channels
with thresholds above the vector particle mass. The HK mass of a vector meson,
even if the most relevant from the point of view of effective
Lagrangians, is somewhat indirect  as it does not correspond
directly to a measured  quantity. It will be seen below that
the HK masses of $\omega$ and $\phi$ mesons are both below
the two--kaon thresholds.

{From} the  rigorous point of view of S--matrix Theory \cite{ELOP},
the relevant mass--shell concept is rather  the pole location
of the propagator. Focusing on the $\omega/\phi$ sector, we have seen 
that the mass squared involves solutions to an eigenvalue problem and 
should satisfy $s-\lambda_\omega(s)=0$ and $s-\lambda_\phi(s)=0$. So the 
pole is certainly located in the complex plane. The issue is basically the same  
for the $\rho$ meson \cite{tony} or the $K^*$'s and is quite general 
\cite{pennington}.
   
There is no reported \cite{PDG98} information on the pole position for the
$\omega$ and $\phi$ mesons. However, using the reported masses ($M_V$) and widths
($\Gamma_V$)\cite{PDG98}, one can state $s_V=M_V^2-i M_V\Gamma_V$ with some
unknown precision. We thus can guess that the poles for the
$\omega$ and $\phi$ mesons are very close to the physical region and,
additionally, that the real part of the $\phi$ pole is very close to the
2--kaon thresholds.

Neglecting its imaginary part compared to its real part, the $\omega$
on--shell pole is certainly much below the 2--kaon thresholds. Therefore,
it is appropriate to state $\delta_\omega(s_\omega) \simeq
\delta_\omega(M_\omega^2)$, and get a real phase.

In the same approximation, the $\phi$ mass is however, slightly above the
2--kaon thresholds and then in a region where $\delta_\phi(s_\phi) \simeq
\delta_\phi(M_\phi^2)$ carries a tiny imaginary part. It is quite instructive to
inspect the loop function $\Pi(s)$ as given in Eq. (\ref{rhoD})  from
Appendix \ref{AA} (with $s_0=4 m_K^2$) and look at its behavior in the
vicinity of the observed $\phi$ mass. The logarithm and
the imaginary part are of order $\simeq 0.05$ GeV$^2$ compared to the
polynomial part ($ \simeq 0.3 $ GeV$^2$). Therefore, it is {\it a priori}
justified to neglect the imaginary part of the loop function while staying
at the $\phi$ mass--shell; somewhat farther beyond  this point, this  statement
would have  surely to be revisited.
 
Therefore, even if somewhat accidental{\footnote {Small widths and/or
small imaginary part for the kaon loop are clearly accidental features.
}}, the matrix $G(\delta_V)$ which gives the $\phi$ and $\omega$
eigenstates is indeed close to a pure rotation matrix when using 
the pole definition for the mass--shell. Anticipating our
results, we have checked that the subtraction polynomial which is fitted
in radiative and leptonic decays does not change this picture. If
one rather defines the mass--shell through the HK masses, $G(\delta_V)$
is a pure rotation matrix.

Additionally,  we have also
neglected in the present approach other $e^2$ contributions like the loops
$\pi^0 \gamma$, $\eta \gamma$, $\eta' \gamma$ which follow from the FKTUY
Lagrangian, as given in Ref.~\cite{rad}. These contribute to provide small
imaginary parts (of order $e^2$) to the self--energies we consider for $s \geq
m_{\pi^0}^2$, $s \geq m_{\eta}^2$, $s \geq m_{\eta'}^2$. Most $VP$ loops
neglected only contribute to self-energies strictly speaking and we shall argue
below why their influence is small. Inspecting the anomalous VVP Lagrangian 
(see Eqs. (A8) or (A14) in Ref. \cite{rad}), the neglected contributions to 
$\Pi_{\omega_I \phi_I}(s)$ are $K^* K$ loops which have HK threshold masses 
above the $\phi$ meson; thus, they would contribute as real quantities in our 
mass region of interest. We shall argue below that such effects are practically 
harmless, essentially because the mixing function is really slowly
varying in the mass region involved in meson decays.

Some of these neglected loops can be computed in closed form (see Appendix 
\ref{AA3}).  We have finally neglected two--loop
effects produced from the FKTUY Lagrangian couplings and double loop
effects produced by the $VPPP$ FKTUY Lagrangian \cite{FKTUY}; this will be 
numerically justified below. Thus, concerning $\Pi_{\omega_I \phi_I}(s)$,
the  most important neglected effects are either two--loops or of
order $e^2$. 

So, whatever the mass--shell definition, when working with on--shell $\omega$ and 
$\phi$ mesons, the transformation from ideal to physical fields is  
very close to a rotation and, then, making this approximation is justified. However, as a 
consequence of VMD at one--loop, $\delta_V(s)$ is $s$--dependent. This is practically 
equivalent  to having {\it two} different mixing angles
$\delta_\omega =\delta_V(m_\omega^2)$, $\delta_\phi =\delta_V(m_\phi^2)$
functionally related. The question to which extent these two
angles differ is addressed below; however, the statistical quality of
the fits in Ref.~\cite{rad} with only one such mixing angle allows to
infer that they are probably close together. On a related topic, let us
recall that a momentum--dependent mixing angle for
pseudoscalar mesons has been recently considered \cite{frere}.

For practical purposes in fit procedures, we have preferred computing the phase 
$\delta_V(s)$ as the HK masses of the $\omega$ and $\phi$ 
mesons{\footnote{Actually, the situation for
the $\phi$ meson is even more  uncertain than sketched above;  within the HLS  model,
and pushing aside electromagnetic interactions,  $K \overline{K}$ scattering
in ($I=0$, $l=1$)--wave is a 2--channel problem with thresholds at $K^0 \overline{K}^0$
and $K^+ K^-$; the domain of definition is a  4--sheeted Riemann surface.
Even if of electromagnetic origin, the distance of the thresholds 
($\simeq 8$ MeV) is not negligible compared to the distance of the PDG mass for the
$\phi$ to the $K^0 \overline{K}^0$ threshold ($\simeq 24$ MeV). Therefore,
branch points and poles are gathered in a tiny neighborhood; this makes
the local topology of the 4--sheeted Riemann surface influencing and, 
at the level of a few MeV's, the physical $\phi$ pole position 
cannot be guessed reliably from its reported mass \cite{PDG98}.}}. 
This sets the $\omega$ mass close to its
PDG value and the $\phi$ mass slightly below the 2--kaon thresholds. As
these HK masses are real, the matrix $G$ is indeed orthogonal. 

When dealing with scattering processes, {\it i.e.} over a broad range for
$s$ extending possibly far above $s=4 m_K^2$, the running of the mixing
angle $\delta_V(s)$ --which then carries an imaginary part--
 is a feature which should play some role. 

\section{Nonet Symmetry Breaking (NSB) in the  HLS Model}
\label{five}

In Ref.~\cite{chpt}, it  was shown that the HLS Lagrangian can undergo
NSB in the PS sector by adding to Eq.(\ref{HLS1}) the following term

\begin{equation}
{\cal L}'= 
\displaystyle \frac{1}{2}[\mu^2\eta_0^2+\lambda \pa_\mu \eta_0 \pa^\mu \eta_0]
\label{HLS10}
\end{equation}
where $\eta_0$ denote the singlet PS field. Such contributions can be
inferred from Chiral Perturbation Theory \cite{leutw,leutwb}  (ChPT).

At the level of radiative decays, the additional kinetic energy term
implies to modify the renormalization condition 
Eq. (\ref{HLS3}) to \cite{chpt} 

\begin{equation}
P=X_A^{-1/2}(P^{ren}_8+xP^{ren}_1)X_A^{-1/2}
\label{HLS11}
\end{equation}
using obvious notations, and by defining the PS NSB parameter
$x=1/\sqrt{1+\lambda}$. Then exact nonet symmetry is defined by $x=1$. In
order to perform both SU(3) breaking and NSB in the PS sector, we refer
here to the change of fields in Eq. (\ref{HLS11}), which diagonalizes the
PS kinetic energy in ${\cal L} + {\cal L}'$ at first order.  It has been
shown that this fits \cite{chpt} perfectly all related information from
ChPT \cite{leutw,leutwb,feldmann} concerning the $\eta/\eta'$ sector at
first order in the breaking parameters. One can then consider quite
reliable the full PS breaking scheme represented by Eq. (\ref{HLS11}).


In the vector sector, the way to introduce NSB is unclear.  
It is even not clear  whether it can be done in full accordance
with the conceptual framework of the HLS model. One could
imagine that an appropriate breaking term of  the HLS  Lagrangian would be 
an additional singlet mass term
\begin{equation}
{\cal L}''= \mu^2_0 (\sqrt{2} \omega_I-\phi_I)^2
\label{HLS12}
\end{equation}
and/or a change of field of the form
\begin{equation}
V= V^{ren}_8+yV^{ren}_1
\label{HLS13}
\end{equation}
which corresponds to having $g$ as coupling constant
for the vector octet matrix and $yg$ for the singlet one.
Departures from $y=1$ would then flag unambiguously NSB in the vector sector.

However, if one focuses on deriving $P\gamma \gamma$ amplitudes from the
VVP Lagrangian, as done in Ref.~\cite{rad}, the difficulty is that this
derivation is impossible -- or at least not obvious -- except if one
assumes that $\mu_0^2$ and $y$ are running and that $\mu_0^2(s=0)=0$
and $y(s=0)=1$. Stated otherwise: the above mentioned derivation
is (trivially) possible only if vector NSB vanishes at the
chiral point. It should be noted that both $X_A$ and $X_V$ breakings on
the one hand \cite{BKY,heath} and the PS NSB breaking \cite{rad} mentioned
above on the other hand, do not require such a condition within the HLS
framework\cite{rad}. However, the
connection between $PV \gamma$ and $PVV$ amplitudes, named Extended VMD in
\cite{FKTUY}, is an extension of the usual VMD assumption which might have
to be reconsidered, if motivated; it is worth mentioning that
Ref.~\cite{FKTUY} does not recommend the possible further extension of the
VMD assumption to box anomalies ($\gamma PPP$).

Considering the HLS model as an effective model, one can, nevertheless,
investigate the question of NSB in the V sector to see whether some
support can be found in experimental data.

A term like Eq. (\ref{HLS12}) results essentially in an additional 
contribution to the phase shift $\delta_V$ and is hard to disentangle from 
the loop effects which naturally follow from the HLS Lagrangian at first non--leading
order. It could only break the relation between the phase shift $\delta_V$
and the loop functions, allowing thus for more freedom in fits.

A breaking term like Eq. (\ref{HLS13})  can be generated (in a presently
unknown way) by breaking the vector Yang--Mills term which is beyond the
scope of the HLS model{\footnote{We mean that some kind of relation should
then exist for $y$, in correspondence with the one relating the PS NSB
parameter $x$ and the (basic) kinetic energy breaking parameter
$\lambda$ recalled above.}}. What is important to note is that, whatever this unknown
procedure, it should  summarize into a relation like Eq.
(\ref{HLS13}) at leading order in breaking parameters.

We limit ourselves to examining the couple of vector NSB
mechanisms given by Eqs. (\ref{HLS12}) and (\ref{HLS13}).

\section{Radiative Decays of Light Mesons}
\label{six}

One can construct axiomatically the decay amplitudes
for the processes $PV \gamma$ assuming $SU(3)\times U(1)$
group structure for PS and V mesons. It has been done
in the classic paper of O'Donnell \cite{odonnel}
(see also the Appendix in \cite{ben2} where some  misprints have been corrected). 
It can be checked easily  that the coupling constants $g_{VP\gamma}$
of \cite{odonnel} can be derived from the $VP\gamma$ term
of the following anomalous Lagrangian
\begin{equation}
{\cal L}_{WZW}= K \epsilon^{\mu \nu \rho \sigma}
{\rm Tr} \left [
\pa_\mu(eQA_\nu+gV_\nu)\pa_\rho(eQA_\sigma+gV_\sigma)P 
\right]
\label{HLS14}
\end{equation}
using Eqs. (\ref{HLS11}) and (\ref{HLS13}) above (with $\ell_A=1$ in order
to limit oneself to NSB only). The coefficient $K= -3/(4 \pi^2 f_\pi)$ is
a normalization fixed by requiring the coupling constant for $\pi^0 \gamma
\gamma$ to be the usual one. Eq. (\ref{HLS14}) gives the connection between
coupling constants for $P \gamma \gamma$, $PV \gamma$ and $PVV$
processes as  expected from the Extended VMD assumption of Ref.~\cite{FKTUY}.

The notations in the present work (or in Ref.~\cite{rad}) and in
Ref.~\cite{odonnel} are connected by $g_{V_8P_8\gamma}=G=-3eg/(8 \pi^2
f_\pi)$, $g_{V_8P_1\gamma}=xG$ and $g_{V_1P_8\gamma}=yG$. Therefore, Eq.
(\ref{HLS13}) holds whatever is the precise underlying relation between
$y$ and more basic (and unknown) vector NSB parameters.

This is quite an interesting pattern. Indeed, if vector NSB is only hidden
inside $\delta_V$, it seems beyond any unambiguous phenomenological
evidence. However, if it affects {\it also} separately the vector coupling
constant by a factor $y$, then departures from $y=1$ can be explored.

Moreover, applying Eq. (\ref{HLS14}) by assuming additionally $\ell_A \ne
1$ is legitimate.  Indeed, as recalled above, Ref.~\cite{chpt} has shown
that the $X_A$ breaking scheme \cite{BKY,heath} is in accord with all
accessible predictions of ChPT \cite{leutw,leutwb,feldmann}.

The coupling constants for radiative decays have been computed starting
from Eq. (\ref{HLS14}) assuming NSB and SU(3) breaking in both sectors.
Additionally, vector meson decays to lepton pairs have been computed
assuming $y \ne 1$; in this case, corrections are quite negligible but
might be considered in order to be complete. All formulae of relevance are
given in Appendix \ref{CC}. In the limit $y=1$, they coincide with the
corresponding expressions derived in Ref.~\cite{rad} starting from the
FKTUY Lagrangian. In the general case where $y$ could differ from unity,
Eq. (\ref{HLS14}) can be used as as alternative way to express the
Extended VMD assumption.

We also work in the framework of the so--called $K^*$ model commented on
below. It introduces an additional breaking procedure \cite{rad} in order
to account for the observed $K^*$ radiative decay rates and a
dimensionless breaking parameter $\ell_T$ fit as \cite{rad} $ \ell_T=1.19
\pm 0.06$.

\section{Comments about the $K^*$ Model}
\label{fiveb}

Referring to \cite{rad} for details, it has been shown that the above
described breaking mechanisms (V and PS NSB's, $X_{A,V}$ breakings)  
altogether account for all leptonic and radiative decays, except for
$K^{*\pm} \ra K^{\pm} \gamma$. More precisely, no way has been found to
allow for the observed ratio of yields $[K^{*0} \ra K^0 \gamma]/[K^{*\pm}
\ra K^\pm \gamma] \simeq 2$; it cannot be else than $\simeq$ 4 within
the approach of O'Donnell \cite{odonnel} or starting from Eq. (\ref{HLS14})
supplemented with the breaking procedures already described.  Quite
interestingly, the non--relativistic quark model (NRQM) \cite{odonnel}
allows for more freedom, depending on the ratio of the quark magnetic
moments $r=\mu_s/\mu_d$

\begin{equation} 
\displaystyle \frac{G_{K^{*0}K^0\gamma}}{G_{K^{*\pm}K^\pm\gamma}}=
-\frac{1+r}{2-r}
\label{morpurgo}
\end{equation}

More recently, motivated by the surprisingly large success of NRQM, G.
Morpurgo has shown that the NRQM predictions $K^*$ radiative decay
coupling constants \cite{GM} are valid in low energy QCD, provided one 
assumes that gluonic contributions are negligible in this energy 
range. This property plays also some role in leptonic decays  \cite{DM}.
Therefore, it is of concern to see how a relation as appropriate as Eq. 
(\ref{morpurgo}) could be derived within the VMD framework.

In order to account for the observed $K^*$ relative rate,
Ref.~\cite{rad} tried first introducing the breaking procedure (named
there $X_W$ breaking)  proposed by Bramon, Grau and Pancheri \cite{BGP}.
This turned out to introduce another breaking matrix $X_W={\rm
Diag}(1,1,1+c_W)$ in the FKTUY Lagrangian \cite{FKTUY}; this can be
symbolically written Tr$[V X_W V P]$.  However, the degree of algebraic
correlations introduced by SU(3) among all single photon radiative decay
modes is such that all fits return $X_W=1$ and do not allow for any
improvement concerning the observed ratio $[K^{*0} \ra K^0
\gamma]/[K^{*\pm} \ra K^\pm \gamma] \simeq 2$.

In order to change this picture, Ref.~\cite{rad} proposed to perform the
replacement $V \Rightarrow X_T V X_T$ in addition to the BGP breaking;  
$X_T$ is assumed to have the classical SU(3) breaking structure $X_T={\rm
Diag}(1,1,1+c_T)$. In this case, one succeeds in fitting  both
observed $K^*$ radiative decay modes. Additionally, it was
phenomenologically observed \cite{rad} that $X_WX_T^4=1$ is well fulfilled
by the full set of radiative decays; this makes all coupling constants
independent of the $X_W-X_T$ breaking, except for the $K^*$ modes which
become

\begin{equation}
\left \{ 
\begin{array}{lll}
\displaystyle G_{K^{*0}K^0\gamma}=&-& G' \displaystyle\frac{\sqrt{\ell_T}}{3}
(1+\frac{1}{\ell_T})\\[0.3cm]
\displaystyle G_{K^{*\pm}K^\pm\gamma}=&~& 
G' \displaystyle\frac{\sqrt{\ell_T}}{3}
(2-\frac{1}{\ell_T})
\end{array}
\right .
\label{kstar}
\end{equation}
where $G'=-3eg/(8 \pi^2 f_K)$ and $\ell_T=(1+c_T)^2$.  The ratio which
can be derived from Eqs. (\ref{kstar}) is in obvious correspondence with
Eq. (\ref{morpurgo}) and allows to identify $r \leftrightarrow 1/\ell_T$.
To our knowledge, the mechanism proposed in Ref.~\cite{rad}, and
just sketched, is the single one able to reproduce the NRQM--Morpurgo relation
for the $K^*$ decay rates in a VMD framework.

The change $V \Rightarrow X_T V X_T$ is in clear correspondence with the
change of (PS) fields imposed by the BKY $X_A$ breaking mechanism (see Eq.
(\ref{HLS3}) above). It resembles what could be a vector field
renormalization, presently lacking within the BKY breaking framework
\cite{BKY,heath}.  If this were so, the $K^*$ sector is the single one
where this effect can be unambiguously visible. Whether it is possible to
derive it rigorously within the HLS framework deserves some effort which
should be supported by new data confirming the reported level for the
$K^{*\pm}$ radiative decay rate.


It is worth mentioning the following remarks
\begin{itemize}
\item The phenomenological observation \cite{rad} $X_WX_T^4=1$ can be interpreted 
if Eq. (\ref{HLS14}) could be rewritten when the  $X_W-X_T$ breaking mechanism
is at work
\begin{equation}
{\cal L'}_{WZW}= K \epsilon^{\mu \nu \rho \sigma}
{\rm Tr} \left\{ [X_T
\pa_\mu(eQA_\nu+gV_\nu)X_T]~X_W~[X_T\pa_\rho(eQA_\sigma+gV_\sigma)X_T]~P 
\right\}
\label{HLS15}
\end{equation}
In order that the $A^2$ term still gives the 2--photon decay amplitudes
predicted by the original WZW Lagrangian \cite{WZW}, $X_WX_T^4=1$ 
becomes  indeed a necessary condition.  

\item If the replacement $V \Rightarrow X_T V X_T$ is indeed the
renormalization condition for the vector fields, then $\ell_V$ in the
standard formulae (given in Appendix \ref{CC} or in Ref.~\cite{rad})
actually hides as many powers of $\ell_T$ as the number of $\phi_I$ fields
the term involves. For leptonic decays $\ell_V$ should then be understood
as $\ell_V \ell_T$. If the above replacement were theoretically motivated,
it would indicate that the $\phi$ HK mass squared is not \cite{BKY,heath}
$m_\rho^2 \ell_V$ but rather $m_\rho^2 \ell_V \ell_T^2$; this enforces our
standpoint that the relevance of any {\it a priori} value for the true
$\ell_V$ is pending and might have to wait for a final answer concerning
vector field renormalization in the HLS--BKY framework.
\end{itemize}

\section{Numerical Analysis}
\label{seven}

In studies published elsewhere \cite{chpt}, it was shown that
the PS mixing angle $\theta_P$ and the PS NSB parameter $x$
were fulfilling 
\begin{equation}
\tan{\theta_P}= \displaystyle \sqrt{2} \frac{1-z}{2+z} x
\label{chpt}
\end{equation} 
exceptionally well ($z=[f_K/f_\pi]^2$). This relation is a consequence of
the small value of the decay constant $F^0_\eta$ recently defined in
ChPT\cite{leutw,leutwb}. Quite interestingly, it projects onto the PS
mixing angle $\theta_P$ most departures from nonet symmetry. From the point
of view of numerical analysis, this also allows to reduce the number of
free parameters by one unit without any change in the fit quality. In all
fits referred to below, this condition has been either relaxed or
requested.  In all cases where fits returned a good probability this
additional requirement has been found to leave the $\chi^2$ unchanged;
setting this condition mechanically improves the probability, as it turns
out to increasing the number of degrees of freedom for exactly the same
$\chi^2$ value. The set of data submitted to fits are the 14 radiative
decays{\footnote{ This counting leaves aside the process $\pi^0 \ra \gamma
\gamma$ which would only fix the value of the decay constant $f_\pi$
already taken from \cite{PDG98} as for $f_K$.}} $VP\gamma$ and $P\gamma
\gamma$ and the 3 leptonic decays $\rho/\omega/\phi \ra e^+e^-$ taken from
the Review of Particle Properties \cite{PDG98}; this represents 17 data
points, {\it i.e.} the largest sample ever submitted successfully to a
fit. All formulae used for fits can be found in Appendix \ref{CC}. We do
not reproduce the reconstructed branching fractions as they are
indistinguishable from the final results in Ref.~\cite{rad} or from those
in Table 2 of Ref.~\cite{chpt}.

In order to explore the question of vector meson mixing, 
we have examined several analysis strategies.

\begin{itemize}
\item[{\bf 1/}] Approximating the right-hand--side of Eq. (\ref{HLS8})
by a constant. This is nothing but the approach developed in Ref.~\cite{rad}
with one constant mixing angle $\delta_V$. In this case we have
either left $y$ free or fixed it to 1. 

\item Setting $y=1$ (no vector NSB), we get $\chi^2/dof=9.14/11$ (61\%
probability) as in Ref.~\cite{chpt} after stating Eq. (\ref{chpt}). From a
statistical point of view, this result can be considered as optimum. The
number of fit parameters is 6, out of which 4 are especially devoted to
the 14 radiative decays ($\delta_V$, $\delta_P$, the universal vector
coupling constant $g$ and the parameter $\ell_T$, specific to the $K^*$
sector) and 2 ($a$, $\ell_V$) concern solely the leptonic sector.
 
\item Leaving $y$ and $\delta_V$ free, we get $\chi^2/dof=8.82/10$ (55 \%
probability); the $\chi^2$ is practically unchanged but the probability is
slightly degraded by having one more free parameter ({\it i.e.} one less
degree of freedom).  In this case we get $y=1.012 \pm 0.022$, a quite
insignificant departure from no vector NSB;  the vector
mixing angle $\theta_V=31.57^{\circ} \pm 0.62^{\circ}$ is still found
below ideal mixing, that is $\delta_V \simeq-3.70^{\circ} \pm
0.62^{\circ}$, quite significantly negative (about a $6~\sigma$ effect).
This result is in agreement with the phase of Dillon and Morpurgo
\cite{DM} which relies on only leptonic decays of vector mesons. Removing
the leptonic decay modes from the fit leaves the $\chi^2$ unchanged
($\chi^2/dof=8.52/10$, 38\% probability) and provides the same values for
$\delta_V$ (in magnitude and sign) and for $y$.

\item Finally, we have forced  $\delta_V =0$ and left $y$ free,
in order to see whether explicit vector NSB could alone do the work
generally attributed \cite{rad,DM} to angular departures from ideal
mixing. In this case departure from nonet symmetry is fit as 
large in the vector sector ($y=0.893\pm 0.005$) as in the PS sector
($x=0.901 \pm 0.017$); however, the fit quality is unacceptably
degraded ($\chi^2/dof=41.81/11$, 2. 10$^{-5}$ probability).

\item[ ] {\bf Comments~:} Concerning the vector mixing angle, several
conclusions follow from  these fits. First, whatever its precise origin,
the effects of angular departures from ideal mixing are fundamental; indeed, whatever
the way used in order  to circumvent it, the gain which can be attributed
to  it is of  the order  of 30 units in the $\chi^2$, a  quite
significant effect for a single parameter. Moreover,
our various fits show that radiative decays and leptonic decays
carry {\it separately} the same information about the magnitude
and sign of $\delta_V$. The same sign information has been reached
by the quite independent approach  of Ref.~\cite{DM} using only
leptonic decays. We come back on this point in Section \ref{nine}.

Explicit departures from nonet symmetry for vector mesons ($y$)  are
statistically insignificant (about $0.5 ~\sigma$) and can clearly be
ignored.  Stated otherwise, if there is nonet symmetry breaking in the
vector sector, it cannot be the manifest U(3) breaking \`a la O'Donnell
\cite{odonnel}, as data sharply favor $g_{V_8P_8\gamma}=g_{V_1P_8\gamma}$.
Quite interestingly, Refs.~\cite{GM,DM} reach a parent conclusion in a
framework quite different from ours; they express it by stating that
gluonic annihilations should have negligible contributions
in light meson decays.

\item[{\bf 2/}] We consider the vector mixing angle $\delta_V$ as the
$s$--dependent function given in Eq. (\ref{HLS8}). In this case, through
the kaon loop, it depends on a well  defined function and
an arbitrary polynomial (see Eq. (\ref{self1})). The behavior of $P(s)$ is
minimally $d_0+d_1s$; however, several attempts led us to go one unit
beyond minimality in subtracting the dispersion relation and then choose
$P(s)=d_0+d_1s +d_2s^2$; the final loop function used is given in Eq.
(\ref{self1}). In Section \ref{four}, we have shown that the appropriate
renormalization condition for the constant term here was $d_0=0$. On the
other hand, analysis of fit results has shown that the renormalization
condition $d_1=0$ is numerically appropriate, despite that 
$d_1$ and $d_2$ happens to be highly correlated (99\%).

Therefore, the function we use for $\delta_V(s)$ depends on one single
free parameter ($d_2$) and on a non--trivial well defined logarithm
function. This single parameter practically generates 2 mixing angles
$\delta_V(m_\omega^2)$ and $\delta_V(m_\phi^2)$ functionally related with
each other. In part {\bf 1} just above, the corresponding free parameter was a
single mixing angle, so the situation is somewhat different. Interestingly,
this functional dependence correlates the HLS parameter $a$ and 
the BKY  breaking parameter $\ell_V$  to  the sector of radiative decays 
(see Eq. (\ref{HLS8b})), which is obviously not the case when
having a single constant vector mixing angle (see {\bf 1/} above).
As we have seen that explicit vector NSB does not provide any improvement, 
we have set $y=1$ in all fits referred to hereafter.

\item We have first performed fits with the full expression in Eq.
(\ref{HLS8b}) computed at the HK masses for $\omega_I$ and $\phi_I$ as
they occur in Eq. (\ref{HLS4}), using at each minimization step the
current values for $a$, $g$ and $\ell_V$. The fit quality reached is
$\chi^2/dof=7.89/11$ (72\% probability). Compared with the (first
mentioned) reference fit in ({\bf 1/}), the gain in probability is not
statistically significant. However, as it is the same data set and the
same number of free parameters, this could point towards some evidence
that departure from ideal mixing is indeed observed mass--dependent. Only
much improved data could allow to go farther.

The fit parameter values are $a=2.44 \pm0.04$, $G=0.703 \pm 0.002$
GeV$^{-1}$ (the relation between $G$ and $g$ is given in Eq.
(\ref{brk1})), the pseudoscalar mixing angle is $\theta_P=-10.30^{\circ}
\pm 0.20^{\circ}$ as always above. The main purpose of Ref.~\cite{chpt}
was to show that this fits perfectly with all ChPT predictions, and
the usual (ChPT) mixing angle $\theta_8$ fulfills $\theta_8 \simeq 2 \theta_P$.
 
The breaking parameter is $\ell_V=1.42 \pm 0.03$ and we get
$d_2=(0.147\pm0.008)~ 10^{-2}$ GeV$^{-2}$. All other values are nearly identical
to the corresponding ones in the fits with one vector mixing angle
\cite{rad}. The NSB $x$ parameter value corresponding
to the PS mixing angle is fit to $x=0.900 \pm 0.017$.

Therefore the varying vector mixing angle is consistent with the full data
set; its values are tightly connected with the very small (but
significantly non--zero) value for $d_2$. Using the parameters above and
their errors Eq. (\ref{HLS8b}) allows to compute $\delta_V(m_\omega^2)=
-2.22 ^{\circ}\pm 0.21^{\circ}$ and $\delta_V(m_\phi^2)= -3.44 ^{\circ}\pm
0.30^{\circ}$. The errors here are of course not independent, they
nevertheless allow to understand why a single mixing angle works so well.
\end{itemize}

\section{Guess Estimate of Neglected Anomalous Loops Effects}

In the numerical analysis above, we have not found any need 
to decouple the mixing angle from the $\Pi_{\omega _I\phi _I}(s)$ loop expression, 
as it could be if there were a significant vector NSB contribution to the mixing angle
function, or if significant departures from Eq.~(\ref{HLS8b}) were actually at work.
It is this last point that we shall comment on further here.
 
Anomalous $\omega_I$ couplings are  different from those for $\phi_I$;
taking into account the major $\omega$ decay mode, these could play
some role in the present problem, as the connection 
in Eq.~(\ref{HLS8b}) between $\Pi_{\omega _I\phi _I}(s)$ and 
$\Pi_{\omega _I \omega _I}(s)-\Pi_{\phi_I \phi _I}(s)$ might have been broken 
in real life. 

Among the neglected effects, $K \overline{K} \pi$ intermediate states
from the VPPP Lagrangian \cite{FKTUY} could affect the content of
the transition amplitude $\Pi_{\omega _I\phi _I}(s)$ and of 
both self--energies  $\Pi_{\omega_I \omega_I}(s)$ and $\Pi_{\phi_I\phi_I}(s)$.
As already stated, this  (double) loop is real in the mass region of interest for 
light meson decays. Then, it would supplement the kaon loop by logarithmic
contributions and not influence the fits significantly, as the subtraction
term is adjusted numerically in any case. 

Other double loop effects might mainly affect the $\omega_I$ and $\phi_I$
self--energies. In numerical analyses, because one fits the (real) subtraction
polynomial, the real parts of these are of little importance for the 
reason sketched just above. Some problems, however, might arise 
because of the imaginary parts.  

Such imaginary parts can follow from 3--pion intermediate states,
generated by either $\omega_I \ra \rho \pi$ followed by $ \rho \ra \pi \pi$,
or by a possible  $\omega_I \ra  \pi \pi \pi$ contact term 
\cite{FKTUY,kuraev,Hollenberg:1992nj}. Both would indeed provide an imaginary part 
to the right--hand--side of Eq. (\ref{HLS8b}). 

We have estimated the corresponding effects in several manners
using numerical methods, as these loops (and their imaginary parts)
cannot be computed in closed form. 
Firstly, by computing for  running mass values the imaginary parts in the two
cases mentioned above ($\rho \pi$ or contact term) using the width expression 
from Ref. \cite{kuraev}.
The real part in the right--hand--side of Eq. (\ref{HLS8b}) being $\simeq 0.10$, 
its imaginary parts  varyies from $0.13 \times 10^{-2}$ to $1.7 \times 10^{-2}$ from the $\omega$
to the $\phi$ mass, when attributing the full 3--pion effect to $\rho \pi$.
Attributing instead this effect fully to an $\omega_I \pi\pi\pi$ contact term,
the corresponding numbers were smaller~:  $0.10 \times 10^{-2}$ to 
$0.8 \times 10^{-2}$.
In performing this  exercise, we were using the full $\omega$ width value,
which clearly gives an upper bound.

Secondly, one can assume likely
that $m_{\phi_I}^2-m_{\omega_I}^2 >> |\Pi_{\omega_I \omega_I}(s)-\Pi_{\phi_I
\phi_I}(s)|$ in the region of meson resonances. Therefore we have redone the
fits by removing the loop contribution in the denominator of the
expression for $\delta_V(s)$ in Eqs.~(\ref{HLS8}) and  (\ref{HLS8b}).
The fit obtained is also quite good ($\chi^2/dof=7.94/11$, 72\%
probability), even if negligibly degraded, and the numerical results 
do not appreciably differ from those already mentioned. 

Therefore the mixing angle is not greatly affected by uncertainties 
in $\omega_I$ and $\phi_I$ self--energies related to the neglected loops,
as long as one is only dealing with (on--shell) meson decays. 
Additionally, departures from rotations are found to be numerically far below
the present experimental sensitivity. However, in scattering processes 
where the $\omega$ and $\phi$ mesons occur, these imaginary part effects 
should play some role, as  these (3--pion) imaginary parts grow
like some power ($\simeq 2$)  of $s$
and pills up with the imaginary part of the kaon loop.

\section{Loop Effects in Mass Values}
\label{eight}

It is not the purpose of the present paper to perform a detailed study of
the contribution of self--energies to observed values of vector meson
masses. However, we can limit ourselves to mentioning the effects of 
kaon loops, as this trivially follows from the above fits.

The effective masses can be approximated by  \cite{tony}
\begin{equation}
m_{V eff.}^2 = m_V^2 + \Pi_{VV}(m_V^2) 
\label{mass1}
\end{equation}
It has been checked numerically that rotations have a negligible effect
here and have not been included.

Using the fit parameter values it is easy to get the following numbers
(units are MeV)
\begin{equation}
\left \{
\begin{array}{lll}
m_{\omega}^{HK} = 814.6\pm 6.6~,~~  &m_{\omega}^{ eff.}= 817.6 \pm 6.6 \\[0.5cm]
m_{\phi}^{HK} = 969.8\pm 14.1~,~~    &m_{\phi}^{ eff.}= 986.1 \pm 14.2 
\end{array} 
\right.
\label{mass2}
\end{equation}

Then the effect of kaon loops is to shift moderately the $\omega$ mass
upwards (by 3 MeV), while the shift is important for the $\phi$ (about 16
MeV, 4 times its width!). In order to be really conclusive the other
(anomalous) couplings should be taken into account, but we see already
that effects of real part of self--energies are qualitatively sufficient to
push the effective mass of the $\phi$ far from its HK value and closer to
the observed peak value. This illustrates our statement that observed
masses might be quite different from the masses in the Lagrangian (we
also refer to \cite{klingl,shin} and to \cite{pichowsky}).

\section{Departures from Ideal Mixing in $\omega$ and $\phi$ Decays}
\label{nine}

The origin of departures of the $\omega/\phi$ system from ideal mixing has
been investigated in this paper by analyzing several mechanisms separately
and together. The benchmark is the set of all radiative decays of light
mesons (14 processes $VP\gamma$ and $P\gamma \gamma$) and leptonic decays
of vector mesons (3 modes).

The central part of the various models is the BKY SU(3) breaking scheme.
Its reliability is obviously a crucial condition.

The breaking scheme \cite{heath,rad} in the PS sector seems
reliable as the connection between the HLS model broken in this way and
expectations from ChPT\cite{leutw,leutwb,feldmann} is well reproduced
\cite{chpt}. On the other hand, the way nonet symmetry is broken in the
vector sector is in accordance with basics of group theory as illustrated
by the correspondence between the model we propose and the standard
formulation of O'Donnell \cite{odonnel}.

The BKY SU(3) breaking \cite{BKY,heath} in the vector sector is harder to
evaluate directly; it results essentially in shifting apart the
$\omega_I$ and the $\phi_I$ HK masses and in a slight modification of the
leptonic decay rates. However, the leptonic widths of the vector mesons
depend on this breaking parameter (named $\ell_V$ above) and also on the
HLS parameter $a$ (see Eqs. (\ref{brk6}));  if this breaking procedure
were not appropriate, one may guess that fits would return a value for $a$
inconsistent with its value fit in other independent data sets. 

However, the present fit with a varying vector mixing angle gives 
$a=2.44 \pm0.04$ (close to the result with a fixed angle: $a=2.50 \pm0.03$ \cite{rad}), 
in better agreement with the value coming out from
fit to the $e^+e^- \ra \pi^+ \pi^-$ world data \cite{barkov} 
\cite{ben0} ($a=2.37 \pm0.02$), or to the most recent (and
independent) data set  \cite{akhmetshin} ($a=2.38 \pm0.02$).
One should remark that a varying  mixing angle
(with $Z$ at its  physical value 2/3) makes leptonic and radiative 
decays providing a value  for $a$ quite consistent  with pion form 
factor studies. 

In the pion form factor, the prominent feature accounted for by the HLS
$a$ parameter is the strength of a direct coupling $\gamma \pi^+ \pi^-$ relative to
the $\rho^0$ contribution; its effect extends from threshold to about 1
GeV. The pion form factor is free of any influence of SU(3) breaking
(noticeably $X_V$) and then the fit value for $a$ is free of any
correlation with $\ell_V$. Therefore, there is also no manifest reason to
suspect that the BKY SU(3) breaking in the vector sector could be
questionable.

In order to be more exhaustive, we have tried including nonet symmetry
breaking in the vector sector in two different ways.  Finally, we have
considered the challenging effect produced by kaon loops in generating
$\omega_I \leftrightarrow \phi_I$ transitions which forces to rotate the
ideal fields in order to diagonalize the vector mass term.

Having performed a crossed study of all possible effects together or
separately, we reached the following conclusions:

\begin{itemize}

\item Whatever its origin,  an angle $\delta_V$ exhausts (by far) the 
best fit quality, and this quality is statistically optimum. 
An {\it explicit} vector NSB ($y$) is unable to produce a 
comparable effect.

\item A mass--dependent phase shift $\delta_V(s)$ behaves as well, and
even somewhat better, without introducing more freedom in the fits. This
could be considered as a slight evidence in favor of an observed
mass--dependence of the mixing angle, in functional accordance with loop
expressions in the HLS model.

\item Leaving free the manifest NSB parameter $y$ cannot mimic the effect
of $\delta_V$ (constant or not). Moreover the value for $y$ returned by
all fits is consistent with no vector NSB.

\item Whatever the context, $\delta_V$ is negative ($\delta_V \simeq
-3^{\circ})$, confirming  within VMD the analysis of Dillon and Morpurgo
\cite{DM}, who share certainly the same conventions as ours. 
When there are effectively two such angles, both are found
negative and close together. For completeness, this sign for $\delta_V$
is tightly connected with our definition of $\phi_I=-s\overline{s}$ and  
with the signs in the matrix $G(\delta_V)$. This corresponds to an
$\omega/\phi$ mixing angle slightly smaller than its ideal value
($\theta_V \simeq 32^\circ$). The fit probabilities are always
of the order 60\% to more than 70\%.

\item We have carefully tried to find secondary acceptable minimum $\chi^2$
solutions. The aim was to look for a solution with somewhat different
parameter values and noticeably a positive value for $\delta_V$
(and, thus, a vector mixing angle greater than $\simeq 35^\circ$).
We never reached a $\chi^2$ better than about 40 units when forcing $\delta_V$ to
stay positive or zero. This means a fit probability of the order
10$^{-5}$. 

\end{itemize}

So, the conclusion about the mixing angle coming out from fits to 
radiative and leptonic decays is stable under a large variety of conditions. 
It is obtained within a highly constrained scheme with very few parameters
and quite good probabilities (above the 60\% level). The data used come from
different kind of experiments and can be widely considered statistically
independent of each other.

Actually, it  is quite trivial to prove, from within the non--anomalous HLS Lagrangian 
alone, that an average $\delta_V$ is surely negative while relying on only leptonic decays.
Our definition being $\phi_I=-s\overline{s}$, it is trivial to
show that the $V-\gamma$ coupling constants (see Eqs. (\ref{brk6}))
of the BKY broken HLS Lagrangian fulfill{\footnote{These relations
correct each for a misprint in unnumbered relations in Ref. \cite{rad} nearby  
Eqs. (22) and (23) (which are instead both correct).}}
\be 
\begin{array}{ll}  
f_{\omega \gamma} \cos{\delta_V} - f_{\phi \gamma} \sin{\delta_V} =
\displaystyle \frac{f_{\rho \gamma}}{3}~~~~, &
 f_{\omega \gamma} \sin{\delta_V}+f_{\phi \gamma}\cos{\delta_V}=
\displaystyle \frac{f_{\rho \gamma}}{3} \sqrt{2}\ell_V
\end{array} 
\label{ideal1} 
\ee
up, possibly,  to higher order terms in vector NSB (see Eqs. (\ref{brk8}))
and without any influence of radiative decay models. If $\phi_I=-s\overline{s}$, 
all three $f_{V\gamma}$ above should be positive \cite{heath}. The first relation
can be considered as  an equation for $\delta_V$ in terms of leptonic decay 
data \cite{PDG98} (units  are GeV$^2$): $f_{\rho \gamma}=0.119 \pm 0.003$,
$f_{\omega \gamma}=(3.586 \pm 0.060)~10^{-2}$ and
$f_{\phi \gamma}=(7.933 \pm 0.114)~10^{-2}$. It is trivial to solve it
and get $\delta_V=-2.79^\circ \pm 0.84^\circ$. 

Therefore, our results merely illustrate that the  sign information
for $\delta_V$,  hidden in radiative decays is in perfect agreement
with the sign which can be obviously exhibited 
in leptonic decays. So, in the HLS approach, the algebraic value 
for $\delta_V$ follows from  the radiative {\it and} from the 
leptonic sectors {\it separately}. 

However, there is also a result by Achasov, Kozhevnikov and 
Shestakov \cite{achasovp} which predicted a correlation between 
the signs of $R= [f_{\phi \gamma} G_{\phi \rho \pi}]/[f_{\omega \gamma} G_{\omega \rho \pi}]$
and the interference pattern in the neighborhood of the $\phi$ meson
in the annihilation processes $e^+ e^- \ra \pi^+ \pi^- \pi^0$.
 {From}  the results reported in \cite{dolinsky}, the conclusion was that $R$  should
be negative and this should imply a positive value for $\delta_V(s)$.
Recent analyses of the $e^+ e^- \ra \pi^+ \pi^- \pi^0$ cross section
by Achasov {\it et al.} \cite{achasov2} seem to confirm their conclusion.
The origin of this disagreement has not been explored and could point 
toward an interesting puzzle.

Indeed, taking into account the stability of our fit results
inside the HLS framework and the cross--check represented by  the independent analysis 
of Dillon and Morpurgo \cite{DM}, we consider our small negative value for 
$\delta_V$ unavoidable when using (even separately) radiative and leptonic decays
within the HLS framework. 
A varying mixing angle leads to the same result.
Possible real part effects of the neglected loops, might be practically
accounted for by the numerical values for $d_1$ and $d_2$. Imaginary part
effects have be shown to be small in the mass region of relevance for
(on--shell) vector meson decays.

\section{Conclusion}
\label{ten}

We have studied in detail the origin of departures of
the $\omega/\phi$ system from ideal mixing in the non--anomalous HLS Model. 
Whatever its origin, the mechanism at work is clearly the existence of 
$\omega_I \leftrightarrow \phi_I$ transitions. The simplest origin
of these transitions are the kaon loops to which $\omega/\phi$ couple.
Because they involve on--shell particles, in resonance decays the $\omega/\phi$ mixing 
occurs essentially through a field rotation, as traditionally assumed. Departures 
from rotation could be presently observable in scattering processes involving 
far off  mass--shell $\omega$ and $\phi$ mesons as internal lines. 

In order to make this usable for phenomenological purposes, we have
defined an effective Lagrangian piece containing all loops of the
non--anomalous HLS Lagrangian only. Summing up the standard broken Lagrangian
with this effective piece confirms that
$\omega_I \leftrightarrow \phi_I$ transitions are inherent to the HLS
model, broken or not, in accordance with the
Schwinger--Dyson equation. The transformation from ideal to physical
fields has been studied.  

A possibly new result is that this mixing angle appears as a
well-defined momentum--dependent  function $\delta_V(s)$, depending also on a
subtraction polynomial $P(s)$. Its constant coefficient ($d_0$) should be
zero in order to recover the $P\gamma \gamma$ decay amplitudes from the
FKTUY Lagrangian. 

We have left aside the consequences for scattering amplitudes and focused
on the radiative and leptonic decays of light mesons. We have thus shown
that the model proposed in Ref.~\cite{rad} was perfectly consistent in
either of its two possible variants: constant or invariant--mass 
dependent $\delta_V$. Data give some slight evidence in favor of this
dependence, however a constant value for $\delta_V$ is, presently, a
good approximation.  Our result compares well with the
independent analysis by Dillon and Morpurgo using a quite different
framework. The value for $a$ is found in good agreement
with independent pion form factor studies;  the agreement is somewhat
better with varying mixing angle than with a fixed one.

$\omega/\phi$  mixing is generated by loop effects,
without any help of symmetry breaking. Following some trend, we have nevertheless
examined whether a successful description of radiative and leptonic decays
could be reached (or improved) by adding nonet symmetry breaking  in
the vector sector of the HLS Lagrangian.  Instead of the
remarkable effect of this kind of breaking in the PS sector, we have found no indication,
statistically significant, that vector NSB could help in a better
understanding of the data {\it as a whole} (we mean the 17 decay modes
considered altogether, as it should). If it exists, vector NSB could however
be hidden inside $\delta_V$ and thus hard to disentangle from genuine loop
effects; these are, however, widely sufficient in order to understand qualitatively
and quantitatively all the data we have examined at their present level of accuracy.

{From} a specific point of view, all the variants explored (vector NSB, fixed
or varying $\delta_V$) converge towards a $\omega/\phi$ mixing angle slightly below
its ideal value for on--shell $\omega$ and $\phi$. Stated otherwise, a constant 
$\delta_V$ is at about
$-3^\circ$; a varying one is equivalent to two such angles (one at the
$\omega$ mass, one at the $\phi$ mass) but both are negative, close
together and also $\simeq -3^\circ$.

\vspace{1.0cm}
\begin{center}
{\bf Acknowledgements}
\end{center}
HOC was supported by the
US Department of Energy under contract
DE--AC03--76SF00515.

\appendix

\section{The $K \overline{K}$  or $\pi \pi$ Loop Expression}
\label{AA}

The loop expressions for a vector particle decaying into two pseudoscalar
mesons of equal masses can be computed by means of dispersion relations
\cite{klingl} or by using Pauli--Villars regulators \cite{shin}. We derive
here this expression without performing any explicit integration, by
relying on properties of analytic functions and on the 
uniqueness properties of analytic continuation.

Let us denote by $V$ a particular vector meson and by $P$ and
$\overline{P}$ the pseudoscalars of the pair to which they couple; the
common mass to the pseudoscalars is denoted $m_P$.  Let us also denote
$\Pi(s)$ the $P\overline{P}$ loop function.

{From} general principles $\Pi(s)$ is  a real analytic function of $s$
( {\it i.e.} fulfilling $ \Pi(s)= \Pi^*(s^*)$), real below the threshold 
located at $s_0=4 m_P^2$. Its imaginary part  above $s_0$ can be computed 
using Cutkotsky rules or by means of the partial width $V \ra P \overline{P}$
(${\rm Im} \Pi(s)=-\sqrt{s} \Gamma(s)$). We thus have
\begin{equation}
{\rm{Im}} \Pi(s)=-\displaystyle \frac{g^2_{VP\overline{P}}}{48 \pi}
\frac{(s-s_0)^{3/2}}{s^{1/2}}
\label{rho1}
\end{equation}

The analytic function $\Pi(s)$ fulfills (at least) a twice subtracted
dispersion relation, as clear from power counting in the expression for
${\rm{Im}} \Pi(s)$ above. This equation can be written

\begin{equation}
\Pi(s)= \displaystyle P(s) +  \frac{s^2}{\pi} \int_{s_0}^{\infty}
\frac{{\rm{Im}}\Pi(z)}{z^2(z-s+i\epsilon)} dz
\label{rho1a}
\end{equation}
exhibiting that the single cut on the physical sheet lies along the
physical region $s \geq s_0$. $P(s)$ denotes a polynomial of (at least)
first degree.  The coefficients in $P(s)$ should be real and fixed by
means of (external) renormalization conditions, as in Chiral Perturbation
Theory (ChPT). As noted above, the minimal degree of $P(s)$ is 1; however,
the actual degree of this polynomial (and, hence, the actual number of
subtractions to the dispersion relation above) depends on the assumed
behavior of $\Pi(s)$ at infinity{\footnote{Depending on this, $P(s)$ could
also be some entire function of $s$, real for real $s$.  All the general
properties listed here follow from the standard analytic S--matrix theory
\cite{ELOP};  they apply obviously to all amplitudes constructed from any
acceptable Lagrangian.}}.  The coefficients of this (arbitrary) polynomial
need then to be fixed using renormalization conditions such as the values
of $\Pi(s)$ and/or its derivatives at some point. This can be chosen as
the point $s=0$, if one likes to connect with ChPT; other ways are
possible which will not be examined here (see Refs.~\cite{klingl,shin}).

On the other hand, the theory of analytic function teaches that, if we can
find {\it one} analytic function $\Pi(s)$ such that Eq. (\ref{rho1a})
holds, then this solution is unique up to a polynomial (or an entire
function) real for real $s$. That is, two arbitrary analytic
solutions to Eq. (\ref{rho1a}) differ only by a polynomial with real coefficients.  
In this section, and in the following ones dealing with loop computations,
we specialize to a minimally subtracted dispersion relation.

Now, let us define the function $K(s)$ by

\begin{equation}
\Pi(s)=\displaystyle 
\frac{g^2_{VP\overline{P}}}{48 \pi} \frac{s-s_0}{s} K(s) +P(s)
\label{rho2}
\end{equation}

\noindent where $P(s)$ is the  polynomial already defined; $K(s)$
is real for real  $s$ below threshold and we have\begin{equation}
\left \{
\begin{array}{ll}
{\rm{Im}} K(s)=-s^{1/2} (s-s_0)^{1/2} ~~~, ~~~~ (s \ge s_0)  &~\\[0.4cm]
K(s)= \displaystyle c_0+c_1 s + c_2 s^2 + \frac{s^3}{\pi} \int_{s_0}^{\infty}
\frac{{\rm{Im}}K(z)}{z^3(z-s+i\epsilon)} dz
& ~~ 
\end{array}
\right.
\label{rho3}
\end{equation}
\noindent The dispersion relation for $K(s)$ should be subtracted three
time in order to remove the constant term at $s=0$ which would produce a
simple pole for $\Pi(s)$ as clear from the first Eq. (\ref{rho2}). This is
the minimum number of subtractions consistent with integrability
conditions.

One can construct easily such  a function,
denoted $\varphi(s)$, real in the interval $0\le s \le s_0$. It is

\begin{equation}
\varphi(s)=-\displaystyle \frac{i}{\pi} s^{1/2}(s_0-s)^{1/2}
\ln \frac{(s_0-s)^{1/2} +i s^{1/2}}{(s_0-s)^{1/2} -i s^{1/2}}
\label{rho4}
\end{equation}

It can be rewritten~:
\begin{equation}
\begin{array}{lll}
\varphi(s)=\displaystyle \frac{2}{\pi}
s^{1/2}(s_0-s)^{1/2}
\arctan{\sqrt{\frac{s}{(s_0-s)}}}~~, & 0 \le s \le s_0
\end{array}
\label{rho5}
\end{equation}

Eq. (\ref{rho4}) can easily be continued above $s_0$ (by winding clockwise
around this point by an angle of $\pi$ radians)
and below $s_c=0$, the crossed threshold  (by winding counter--clockwise
by an angle of $\pi$ radians). This gives the function $K(s)$ 
on the whole real axis.
The  constants $c_i$ in rels. (\ref{rho3}) are fixed by requiring that

\begin{equation}
\lim_{s \rightarrow 0}K(s) = 0  ~~~ ,~~~\lim_{s \rightarrow 0}
\displaystyle \frac{d}{ds}K(s) = 0  ~~~
{\rm and}~~~\lim_{s \rightarrow 0} 
\displaystyle \frac{d^2}{ds^2}K(s) = 0.
\label{rhoB}
\end{equation}

Then,  assuming the minimal number of
subtractions, the general solution for $\Pi(s)$ is, for real $s$

\begin{equation}
\left \{
\begin{array}{lll}
~~& \Pi(s)=d_0+d_1s + Q(s) \\[0.4cm]
~~& Q(s)=\displaystyle \frac{g^2_{VP\overline{P}}}{24 \pi^2}
\left[ G(s) +s_0 -\frac{4}{3}s\right]
 \\[0.4cm]

s \le 0~~~~: & G(s)= ~\displaystyle \frac{1}{2}\frac{(s_0-s)^{3/2}}{(- s)^{1/2}}
\ln \frac{(s_0-s)^{1/2} - (-s)^{1/2}}{(s_0-s)^{1/2} + (-s)^{1/2}} \\[0.4cm]
0 \le s \le s_0~~: &
G(s)=-\displaystyle \frac{(s_0-s)^{3/2}}{s^{1/2}}
\arctan{\sqrt{\frac{s}{(s_0-s)}}} \\[0.4cm]
s \ge s_0 ~~~:& G(s)=-\displaystyle \frac{1}{2} \frac{(s-s_0)^{3/2}}{s^{1/2}}
\left[ \ln \frac{s^{1/2}-(s-s_0)^{1/2}}{s^{1/2} + (s-s_0)^{1/2}}\right]\\[0.4cm]
~~ & -\displaystyle \frac{i \pi}{2} \displaystyle \frac{(s-s_0)^{3/2}}{s^{1/2}}\\[0.4cm]
\end{array}
\right.
\label{rhoD}
\end{equation}

The behavior of $\Pi(s)$ near $s=0$ is simply $d_0+d_1 s+ {\cal{O}}(s^2)$,
where $Q(s)$ behaves like ${\cal{O}}(s^2)$ near the origin. This result
coincides with the one of Ref.~\cite{klingl}. By performing more
subtractions, one could also choose to fix externally the $s^2$ behavior
of the loop near the origin, etc \ldots

The results here apply directly to loops like $\pi\pi$ or $K \overline{K}$
and to the $\rho$, $\omega$ and $\phi$ self--energies with an appropriate
choice of the specific $VP\overline{P}$ coupling constant.
 
\section{The $K \pi$ Loop Expression}
\label{AA2}
 
It is not of common custom to give the loop expression for
vector particles coupling to a pair of unequal mass PS mesons 
(see however \cite{todorov}).
Its derivation  is tightly connected
with the previous case. We limit ourselves to the minimally subtracted
case, as before; it is trivial (and tedious) to go beyond.

In order to fix notations, we identify
this case with $K^* \ra K \pi$. The imaginary
part of the loop is

\begin{equation}
{\rm Im} \Pi(s) = - \displaystyle \frac{g^2_{K^*K\pi}}{24 \pi} \frac{(s-s_0-s_c)}{\sqrt{s}} p
\label{kstar1}
\end{equation}
where we have defined $s_0=(m_K+m_{\pi})^2$ and $s_c=(m_K-m_{\pi})^2$, the
direct and crossed thresholds. $p$ is the cms momentum of the decay
products $p=1/2\sqrt{(s-s_0)(s-s_c)/s}$. Let us also define, as
previously, the function $K(s)$ and the subtraction polynomial $P(s)$ by

\begin{equation}
\Pi(s) = \displaystyle \frac{g_{K^*K\pi}^2}{48 \pi} \frac{s-(s_0+s_c)}{s}K(s) + P(s)
\label{kstar2}
\end{equation}
where the function $K(s)$ obeys the three time subtracted dispersion
of Eq. (\ref{rho3}) with 

\begin{equation}
\begin{array}{ll}
{\rm{Im}} K(s)=-(s-s_0)^{1/2}(s-s_c)^{1/2} ~~~, & s \ge s_0 
\end{array}
\label{kstar3}
\end{equation}

The subtraction constants $c_i$ are chosen such as $K(s) \simeq {\cal O}(s^3)$ 
when $s \ra 0$. The function $K(s)$ can be easily constructed as before
between $s_0$ and $s_c$, and continued below $s_c$  and  above $s_0$.
Let us now define

\begin{equation}
\left \{
\begin{array}{ll}
A=s_0+s_c  -\displaystyle 
\frac{1}{2\sqrt{s_0s_c}} \left[ (s_0+s_c)^2+2s_0s_c \right]
\ln{\frac{m_{\pi}}{m_K}}
 \\[0.5cm]
B=\displaystyle -\left[
\frac{(s_0+s_c)^2+4s_0s_c}{4s_0s_c}\right]
-\frac{1}{8} \frac{s_0+s_c}{(s_0s_c)^{3/2}}\left[
(s_0-s_c)^2-2s_0s_c
\right]\ln{\frac{m_{\pi}}{m_K}}
\end{array}
\right.
\label{kstar5}
\end{equation}

Then, for real $s$, the final solution for $\Pi(s)$ is given by~:

\begin{equation}
\left \{
\begin{array}{lll}
~~& \Pi(s)=d_0+d_1s +Q(s) \\[0.4cm]
~~& Q(s)=\displaystyle \frac{g^2_{K^* K \pi }}{48 \pi^2}
\left[ G(s) +A +B s \right]
 \\[0.4cm]

s \le s_c~~~~: & G(s)=\displaystyle
(s-(s_0+s_c))\frac{(s_0-s)^{1/2}(s_c-s)^{1/2}}{s}\ln \left[
\frac{(s_0-s)^{1/2}-(s_c-s)^{1/2}}{(s_0-s)^{1/2}
+(s_c-s)^{1/2}} \right] \\[0.4cm]
~~&+\displaystyle \frac{\sqrt{(s_0s_c)}(s_0+s_c)}{s} 
\ln{ \frac{m_{\pi}}{m_K}}
 \\[0.4cm]
s_c \le s \le s_0~~: &
G(s)=\displaystyle
2 (s-(s_0+s_c))\frac{(s_0-s)^{1/2}(s-s_c)^{1/2}}{s}
\arctan{\left[\frac{s-s_c}{s_0-s}\right]^{1/2}} \\[0.4cm]
~~&+\displaystyle \frac{\sqrt{(s_0s_c)}(s_0+s_c)}{s} \ln{ \frac{m_{\pi}}{m_K}}
  \\[0.4cm]
s \ge s_0 ~~~:& G(s)=-\displaystyle (s-(s_0+s_c))
\frac{(s-s_0)^{1/2}(s-s_c)^{1/2}}{s}\ln \left[
\frac{(s-s_c)^{1/2}-(s-s_0)^{1/2}}{(s-s_c)^{1/2}
+(s-s_0)^{1/2}} \right]\\[0.4cm]
~~&+\displaystyle \frac{\sqrt{(s_0s_c)}(s_0+s_c)}{s} 
\ln{ \frac{m_{\pi}}{m_K}}\\[0.4cm]
 ~~&-i \displaystyle \pi \frac{(s-s_0)^{1/2}(s-s_c)^{1/2}}{s}(s-(s_0+s_c))
 \\[0.4cm]
\end{array}
\right.
\label{kstar6}
\end{equation}
where the function $Q(s)$ behaves like $s^2$ near the chiral point.
This result is a non trivial  extension of the previous case. Going
to other PS meson pairs is easily performed by changing $m_\pi$ and
$m_K$ by resp. the lightest and heaviest meson mass.

\section{Some Neglected Loops}
\label{AA3}

It could be useful for future developments including anomalous contributions,
or for other purposes, to have at disposal $V P$ and $\gamma P$ loop expressions 
which can play some (presently minor)  role in estimating self--energies. Double loop 
expressions $PPP$ cannot be computed in closed form; they might also contribute little to
meson self--energies \cite{pichowsky}.   

\subsection{$V P$ Loops}

Their main effect, at the present level of accuracy of the data,
could be the contribution for $\rho \pi$ to $\omega_I$ self--energy.
It is presently overwhelmed by the HK mass values. Other possible 
contributions to self--energies or transition amplitudes
would involve an  intermediate $\eta$ or $\eta'$ mesons, which
push  the  thresholds quite high compared to the $\phi$ mass.
The masses for vector mesons here should be the HK masses.
We skip detailed proofs, as they follow closely the lines in the 
two Sections above.  

Let us fix notations by focusing on $\rho \ra \omega \pi$.
Stating $s_c=(m_{\omega}-m_{\pi})^2$ and $s_0=(m_{\omega}+m_{\pi})^2$.
We have
\begin{equation}
{\rm{Im}}\Pi(s)=-\frac{g_{\omega \rho \pi}^2}{96 \pi}
\frac{(s-s_0)^{3/2}(s-s_c)^{3/2}}{s}
\label{loop6}
\end{equation}
and define the function $K(s)$ by
\begin{equation}
\Pi(s)=\frac{g_{\omega \rho \pi}^2}{96 \pi} \frac{(s-s_0)(s-s_c)}{s} K(s) +P(s)
\label{loop7}
\end{equation}

Here $\Pi(s)$ obeys a three time subtracted dispersion relation, as
obvious from power counting in Eq. (\ref{loop6}),
and then $P(s)$ is (at least) of degree 2 with arbitrary coefficients
to be fixed by renormalization conditions. 
$K(s)$ obeys the dispersion relation in Eq. (\ref{rho3}) above with
\begin{equation}
{\rm{Im}}K(s)=-(s-s_0)^{1/2}(s-s_c)^{1/2}
\label{loop8}
\end{equation}
The  procedure above  applies and the final solution for $\Pi(s)$ on the real axis, 
with the minimum number of subtractions is
\begin{equation}
\left \{
\begin{array}{lll}
~~& \Pi(s)=d_0+d_1s + d_2 s^2+ Q(s) \\[0.4cm]
~~& Q(s)=\displaystyle \frac{g^2_{\omega \rho \pi }}{96 \pi^2}
\left[ G(s) +A +B s +C s^2\right]
 \\[0.4cm]
s \le s_c~~~~: & G(s)=\displaystyle
\frac{(s_0-s)^{3/2}(s_c-s)^{3/2}}{s}\ln \left[
\frac{(s_0-s)^{1/2}-(s_c-s)^{1/2}}{(s_0-s)^{1/2}+(s_c-s)^{1/2}} \right] \\[0.4cm]
~~&-\displaystyle \frac{(s_0s_c)^{3/2}}{s}\ln{\frac{m_{\pi}}{m_{\omega}}}
 \\[0.4cm]
s_c \le s \le s_0~~: &
G(s)=\displaystyle
-2 \frac{(s_0-s)^{3/2}(s-s_c)^{3/2}}{s}
\arctan{\left[\frac{s-s_c}{s_0-s}\right]^{1/2}} \\[0.4cm]
~~&-\displaystyle \frac{(s_0s_c)^{3/2}}{s}\ln{\frac{m_{\pi}}{m_{\omega}}}
 \\[0.4cm]
s \ge s_0 ~~~:& G(s)=-\displaystyle
\frac{(s-s_0)^{3/2}(s-s_c)^{3/2}}{s}\ln \left[
\frac{(s-s_c)^{1/2}-(s-s_0)^{1/2}}{(s-s_c)^{1/2}+(s-s_0)^{1/2}} \right]\\[0.4cm]
~~&-\displaystyle \frac{(s_0s_c)^{3/2}}{s}\ln{\frac{m_{\pi}}{m_{\omega}}}\\[0.4cm]
~~&-i \displaystyle \pi \frac{(s-s_0)^{3/2}(s-s_c)^{3/2}}{s}
 \\[0.4cm]
\end{array}
\right.
\label{intxx}
\end{equation}
with the minimum number of subtractions. We have defined
\begin{equation}
\left \{
\begin{array}{ll}
A=s_0s_c \left[ -1 +\displaystyle \frac{3}{2} \frac{s_0 + s_c}{\sqrt{s_0s_c}} \ln{\frac{m_{\pi}}{m_{\omega}}}
\right ] \\[0.5cm]
B=\displaystyle \frac{5}{4} (s_0+s_c) -\frac{3}{8} \frac{(s_0+s_c)^2 + 4 s_0 s_c}{\sqrt{s_0s_c}}
\ln{ \frac{m_{\pi}}{m_{\omega}}} \\[0.5cm]
C=-\left[\displaystyle  \frac{(s_0+s_c)^2}{8s_0s_c}+\frac{4}{3}+
\frac{s_0^3+s_c^3-9(s_0+s_c)s_0s_c}{16(s_0s_c)^{3/2}}\ln{\frac{m_{\pi}}{m_{\omega}}}
 \right ] 
\end{array}
\right.
\label{int11}
\end{equation}
Going from $\omega_I \pi$ loops to other $VP$ loops is straightforward; let us only note
that the formulae above apply  provided the mass ratio argument of the logarithm is chosen smaller 
than 1.

\subsection{$P \gamma$ Loops}

This contribution could be, for some applications, less academic than the previous
one. It is additionally a singular limit of the $VP$ case ($s_c \ra s_0$). Let us specialize
to $\omega \rightarrow \pi \gamma$. The crossed and direct threshold now
coincide at $s_0=m_\pi^2$.  We have
\begin{equation}
{\rm{Im}}\Pi(s)=-\frac{g_{\omega  \pi\gamma}^2}{96 \pi}
\frac{(s-s_0)^3}{s}
\label{photon1}
\end{equation}
which  follow continuously from the case just above. We still define
the  function $K(s)$ by
\begin{equation}
\Pi(s)=\displaystyle \frac{g_{\omega  \pi\gamma}^2}{96 \pi} \frac{(s-s_0)^3}{s}
K(s)+P(s)
\label{photon2}
\end{equation}
which fulfills the dispersion relation Eq. (\ref{rho3}) with
\begin{equation}
\begin{array}{ll}
{\rm{Im}} K(s)=-1 ~~~, & s \ge s_0  
\end{array}
\label{photon3}
\end{equation}
A specific solution to this equation below threshold is the  function
\begin{equation}
\phi(s)=\frac{1}{\pi} \ln{\frac{(s_0-s)}{s_0}}
\label{photon4}
\end{equation}
real for $s \leq s_0$ and 
which is analytically continued above $s_0$ to
\begin{equation}
\phi(s)=\frac{1}{\pi}\ln{\frac{(s-s_0)}{s_0}}-i
\label{photon5}
\end{equation}
Then the solution for $\Pi(s)$ is
\begin{equation}
\left \{
\begin{array}{llll}
~~~~& \Pi(s)= \displaystyle d_0+d_1 s + d_2 s^2+ Q(s) \\[0.6cm]
~~~& Q(s)= \displaystyle \frac{g_{\omega  \pi \gamma}^2}{96 \pi^2}
[G(s)-s_0^2+\frac{5}{2} s_0 s-\frac{11}{6} s^2 ]
\\[0.6cm]
s\le s_0 ~~: & \displaystyle G(s)=-\frac{(s_0-s)^3}{s}\ln{\frac{(s_0-s)}{s_0}} \\[0.6cm]
s \ge s_0~~: & \displaystyle G(s)=\frac{(s-s_0)^3}{s}\ln{\frac{(s-s_0)}{s_0}}
-i \pi\frac{(s-s_0)^3}{s} \\[0.6cm]
\end{array}
\right.
\label{photon8}
\end{equation}
One should note the disappearance of the square root branch point compared to 
the  $VP$  case. Therefore, the collapse of $s_0$ and $s_c$ pulls the logarithmic
branch point, originally far inside the unphysical sheet, to the threshold.
 
\section{Self--Energies and Transition Amplitudes}
\label{BB}

For the purpose of the present work, we consider all loops computed in the
first two Appendices above amputated from their couplings constants (stated
otherwise, all expression there are considered with
unit coupling constants). Now let us denote $\Pi(s)$ the kaon loop and
$\Pi'(s)$ the pion loop. Let us also denote $\Pi''(s,\pi/\eta/\eta')$ the
loops functions for the three PS meson pairs $K-(\pi/\eta/\eta')$.

In exact SU(2) limit, the self--energies for charged and neutral $\rho$
mesons are the same

\begin{equation}
\Pi_{\rho \rho}(s)= 
2 g_{\rho K \overline{K}}^2 \Pi(s)+g_{\rho \pi \pi}^2 \Pi'(s)
\label{other1}
\end{equation}

Likewise, in the same exact SU(2) limit, for all $K^*$ mesons the self--energies are
\begin{equation}
\Pi_{K^* K^*}(s)= 3 g_{K^* K \pi}^2 \Pi''(s,\pi)
+g_{K^* K \eta}^2 \Pi''(s,\eta)+g_{K^* K \eta'}^2 \Pi''(s,\eta')
\label{other2}
\end{equation}

The coupling constants involved in Eqs. (\ref{other1}) and (\ref{other2})
can be read off the Lagrangian given in Ref.~\cite{heath} and are not
explicited here. There is no more contributions involving $\rho$ or $K^*$
mesons, as soon as one neglects terms of order $e^2$ and all anomalous terms. 
There is also no transition from $\rho$ (or $K^*$) meson to any other 
meson in the same approximation.

Let us redefine the generic kaon loop $\Pi(s)$ as in Eq. (\ref{rhoD}) by
going one unit beyond the minimal subtraction scheme for reasons 
explained in the body of the paper. We have
 
\begin{equation}
\left \{
\begin{array}{lll}
\Pi(s) &=&  P(s) + \displaystyle \frac{1}{24 \pi^2}
\left[ G(s) +s_0 -\frac{4}{3}s +\frac{1}{5} \frac{s^2}{s_0} \right]
 \\[0.4cm]\\[0.3cm]
 P(s)&=&  \displaystyle d_0+d_1s + d_2 s^2
\end{array}
\right.
\label{self1}
\end{equation}
where $s_0=4m_K^2$ and $G(s)$  is given by Eq (\ref{rhoD}). Correspondingly
to introducing $d_2 s^2$ we have subtracted the $s^2$ behavior in $G(s)$.

We give now the functions $\Pi_{\omega_I\omega_I}(s)$, $\Pi_{\phi_I\phi_I}(s)$ and
$\Pi_{\omega_I\phi_I}(s)$ which come at next--to--leading order in influencing
the non--anomalous HLS Lagrangian mass term. The bare Lagrangian piece in
Eq. (\ref{HLS5}) implies that the kaon loop in each self--energy comes in
both their charged and neutral modes; they both coincide with $\Pi(s)$
defined above, when assuming exact SU(2) symmetry.  Then, we have

\begin{equation}
\left \{
\begin{array}{lll}
\Pi_{\omega_I\omega_I}(s) &=& 
\displaystyle 2 g_{\omega_I K \overline{K}}^2 \Pi(s) \\[0.5cm]
\Pi_{\omega_I\omega_I}(s) &=& 
\displaystyle 2 g_{\phi_I K \overline{K}}^2 \Pi(s)   \\[0.5cm]
\Pi_{\omega_I\phi_I}(s) &=& 
\displaystyle 2 g_{\phi_I K \overline{K}} g_{\omega_I K \overline{K}}\Pi(s)   
\end{array}
\right.
\label{self2}
\end{equation}
where the factor of 2 accounts for the two kaon loops (charged or neutral
modes). All dependences in the coupling constants $VK \overline{K}$ are
explicit. The coupling constants in Eq. (\ref{self2}) can be read off Eq.
(\ref{HLS5}). One should finally note that we assume the same subtraction
polynomial $P(s)=d_0+d_1s + d_2 s^2$ in all three functions.  This is
highly constraining and might be somewhat relaxed. Indeed, one could
consider separately dispersion relations for the functions
$\Pi_{\omega_I\omega_I}(s)$, $\Pi_{\phi_I\phi_I}(s)$ and $\Pi_{\omega_I\phi_I}(s)$
which could then undergo different renormalization conditions. This turns
out to remark that a condition like ${\rm Im} \Pi_{\omega_I\omega_I}(s)=\lambda
{\rm Im} \Pi_{\phi_I\phi_I}(s)$ ($\lambda $ being a real constant) for $s \geq s_0$
does imply $\Pi_{\omega_I\omega_I}(s)=\lambda \Pi_{\phi_I\phi_I}(s)$, {\it up to
a polynomial with real coefficients}. We have nevertheless
preferred considering as basic the dispersion relation for the elementary
$K\overline{K}$ loop.  Moreover, contributions of order
$e^2$ and other anomalous  contributions are neglected.

\section{Coupling Constants for Radiative Decays}
\label{CC}

Starting from the Lagrangian in Eq. (\ref{HLS14}), and using the breaking 
procedure defined by Eqs.(\ref{HLS11}) and (\ref{HLS13}), one can compute
the coupling constants for all radiative and leptonic decays of relevance. 
Let us define

\begin{equation}
G=\displaystyle 
-\frac{3eg}{8 \pi^2 f_\pi}~~~,~~~ 
G'=\displaystyle -\frac{3eg}{8 \pi^2 f_K}~~~, ~~~Z= [f_\pi/f_K]^2
\label{brk1}
\end{equation}

We shall use in all formulae a single vector mixing angle $\delta_V$. In
the framework where it is approximated by a constant phase (like in
Ref.~\cite{rad}), it is the common mixing angle which expresses departures
from ideal mixing. In the case when one considers a mass dependent mixing,
one has to make the replacement $\delta_V \Rightarrow
\delta_V(s=m_\omega^2)$ in all expressions for $g_{P \omega \gamma}$,
while the replacement is $\delta_V \Rightarrow \delta_V(s=m_\phi^2)$ in
all expressions for $g_{P \phi \gamma}$. In this case, we practically have
{\it two} different vector mixing angles $\delta_V^\omega$ and
$\delta_V^\phi$ functionally related.

Let us define the parameters $h_i$ $(i=1, \cdots ~4)$ which contain all
information about breaking nonet symmetry in the PS and V sectors, while
breaking SU(3) itself only in the PS sector. Indeed, as the BKY breaking
mechanism does not result in a redefinition of the vector fields, the
radiative decays are not sensitive to SU(3) breaking in the vector sector;
additionally, the condition $X_W X_T^4=1$ removes all dependences on $\ell_T$
for all couplings constants except for $K^*$.  Thus, we have

\begin{equation}
\left \{
\begin{array}{lll}
h_1&=& \displaystyle \frac{(1+2y)(1-x)+2(y-1)(2+x)Z}{3} \\[0.5cm]
h_2&=& \displaystyle \frac{(1+2y)(1+2x)+4(y-1)(1-x)Z}{3} \\[0.5cm]
h_3&=& \displaystyle \frac{(y-1)(1-x)+(2+y)(2+x)Z}{3} \\[0.5cm]
h_4&=& \displaystyle \frac{(y-1)(1+2x)+2(2+y)(1-x)Z}{3}
\end{array}
\right .
\label{brk2}
\end{equation}

The full U(3) symmetry limit is $x=y=Z=1$. 
The $VP\gamma$ coupling constants are
\begin{equation}
\left \{
\begin{array}{lll}
G_{\rho^0 \pi^0 \gamma}=& &\displaystyle \frac{1}{3} G \\[0.3cm]
G_{\rho^{\pm} \pi^{\pm} \gamma}=& &\displaystyle \frac{1}{3} G \\[0.3cm]
G_{K^{*0} K^0 \gamma}=&- & 
\displaystyle \frac{G'}{3} \sqrt{\ell_T} (1+\frac{1}{\ell_T})  \\[0.3cm]
G_{K^{*\pm} K^{\pm} \gamma}=& &
\displaystyle \frac{G'}{3} \sqrt{\ell_T} (2-\frac{1}{\ell_T})  \\[0.3cm]
G_{\rho^0 \eta \gamma}=& &\displaystyle \frac{1}{3} G
\left[\sqrt{2}(1-x)\cos{\delta_P}-(2x+1)\sin{\delta_P}\right]
\\[0.3cm]
G_{\rho^0 \eta' \gamma}=& &\displaystyle \frac{1}{3} G
\left[\sqrt{2}(1-x)\sin{\delta_P}+(2x+1)\cos{\delta_P}\right]
\\[0.3cm]
G_{\omega \pi^0 \gamma}=& & \displaystyle \frac{1}{3}G
\left[(1+2y)\cos{\delta_V} -\sqrt{2}(y-1) \sin{\delta_V}\right]\\[0.3cm]
G_{\phi \pi^0 \gamma}=&- & \displaystyle \frac{1}{3}G 
\left[(1+2y)\sin{\delta_V}+\sqrt{2}(y-1)\cos{\delta_V}\right]
\end{array}
\right .
\label{brk3}
\end{equation}

 The other single photon radiative modes provide
\begin{equation}
\left \{
\begin{array}{lll}
G_{\omega \eta \gamma}=&  \displaystyle \frac{1}{9} G 
\left [\sqrt{2} h_1\cos{\delta_V}\cos{\delta_P}
-h_2\cos{\delta_V}\sin{\delta_P} 
-2h_3\sin{\delta_V}\cos{\delta_P} 
+\sqrt{2}h_4\sin{\delta_V}\sin{\delta_P}
\right ]\\[0.3cm]

G_{\omega \eta' \gamma}=&\displaystyle \frac{1}{9} G \left [
h_2\cos{\delta_V}\cos{\delta_P}
+\sqrt{2}h_1\cos{\delta_V}\sin{\delta_P}
-\sqrt{2}h_4\sin{\delta_V}\cos{\delta_P}
-2h_3\sin{\delta_V}\sin{\delta_P}
\right ]\\[0.3cm]

G_{\phi \eta \gamma}=&\displaystyle \frac{1}{9} G \left [
-2h_3\cos{\delta_V}\cos{\delta_P}
+\sqrt{2}h_4\cos{\delta_V}\sin{\delta_P}
-\sqrt{2}h_1\sin{\delta_V}\cos{\delta_P}
+h_2\sin{\delta_V}\sin{\delta_P}
\right ]\\[0.3cm]

G_{\phi \eta' \gamma}=&\displaystyle \frac{1}{9}G \left [
-\sqrt{2}h_4\cos{\delta_V}\cos{\delta_P}
-2h_3\cos{\delta_V}\sin{\delta_P}
-h_2\sin{\delta_V}\cos{\delta_P}
-\sqrt{2}h_1\sin{\delta_V}\sin{\delta_P}
\right ]
\end{array}
\right .
\label{brk4}
\end{equation}

The 2--photon decay modes keep exactly their form as in Ref.~\cite{rad}
($Z=[f_\pi/f_K]^2$)
\begin{equation}
\left \{
\begin{array}{lll}
G_{\eta \gamma \gamma} = && 
-\displaystyle \frac{\alpha_{em}}{\pi \sqrt{3} f_{\pi}}
\left [ \frac{5-2Z}{3}\cos{\theta_P}-\sqrt{2} 
\frac{5+Z}{3}x \sin{\theta_P} \right ]\\[0.3cm] 
G_{\eta' \gamma \gamma} = && 
-\displaystyle \frac{\alpha_{em}}{\pi \sqrt{3} f_{\pi}}
\left [ \frac{5-2Z}{3}\sin{\theta_P} 
+ \sqrt{2} \frac{5+Z}{3}x \cos{\theta_P} \right ]\\[0.3cm] 
G_{\pi^0 \gamma \gamma} = && -\displaystyle  \frac{\alpha_{em}}{\pi  f_{\pi}}
\end{array}
\right .
\label{brk5}
\end{equation}

Finally, the  $V-\gamma$ couplings become
\begin{equation}
\left \{
\begin{array}{lll}
f_{\rho \gamma} = & \displaystyle a f_{\pi}^2 g \\[0.3cm] 
f_{\omega \gamma} = & \displaystyle \frac{f_{\rho \gamma}}{3}
\left [h_5 \cos{\delta_V}+h_6\sin{\delta_V}\right ]\\[0.3cm] 
f_{\phi\gamma} = & \displaystyle  \frac{f_{\rho \gamma}}{3}
\left [ -h_5\sin{\delta_V}+ h_6\cos{\delta_V}\right ]
\end{array}
\right.
\label{brk6}
\end{equation}
where we have defined
\begin{equation}
\left \{
\begin{array}{lll}
h_5=&\displaystyle 1+ \frac{2}{3} (y-1)(\ell_V-1)\\[0.3cm] 
h_6=&\displaystyle\sqrt{2} \ell_V +\frac{2}{3}(y-1)(\ell_V-1)
\end{array}
\right.
\label{brk8}
\end{equation}

The $X_V$ breaking parameter $\ell_V$ has been defined in the main text.
All relations between the coupling constants here and decay rates are
exactly as defined in Ref.~\cite{rad}. One can check that in the limit $y
\ra 1$, the present results coincide with those given in this reference.
Finally, keeping only U(3) breaking to SU(3)$\times$U(1) in both V and PS
sectors, one can also check that the $VP\gamma$ coupling constants here
coincide with the axiomatic results of O'Donnell \cite{odonnel}.

It should  be noted, from Eqs. (\ref{brk8}), that NSB in the vector sector
for leptonic decays undergo a further suppression  by SU(3) breaking. Finally,
as noted in the  main text, if the  $X_T$ breaking has to be understood as
the relevant vector field renormalization, $\ell_V$ should be understood
as $ \ell_V \ell_T$ without further changes in the other relations above.

\end{document}